\documentclass[fleqn,usenatbib]{mnras}
\usepackage[T1]{fontenc}
\usepackage{ae,aecompl}
\usepackage{graphicx}
\usepackage{natbib}
\pdfoutput=1
\usepackage{amsmath}
\usepackage{amssymb}
\usepackage{threeparttable}
\usepackage{float}
\usepackage{epstopdf}
\usepackage{newtxtext,newtxmath}

\usepackage{etoolbox}
\makeatletter

\let\oldAA\AA
\renewcommand{\AA}{\text{\normalfont\oldAA}}

\newcommand{\lyas}{Ly$\alpha$\ }

\newcommand{\fesc}{$f_\mathrm{esc}$}
\newcommand{\fescs}{$f_\mathrm{esc}$\ }

\newcommand{\HeIIL}{\hbox{{\rm He}\kern 0.1em{\sc ii}\kern 0.1em{$\lambda1640$}}}
\newcommand{\HeII}{\hbox{{\rm He}\kern 0.1em{\sc ii}}}
\newcommand{\HeIIl}{\hbox{{\rm He}\kern 0.1em{\sc ii}\kern 0.1em{$\lambda1640$}}}
\newcommand{\CIIIL}{\hbox{{\rm C}\kern 0.1em{\sc iii}]\kern 0.1em{$\lambda1907,\lambda1909$}}}
\newcommand{\CIVL}{\hbox{{\rm C}\kern 0.1em{\sc iv}\kern 0.1em{$\lambda1548,\lambda1550$}}}
\newcommand{\CIII}{\hbox{{\rm C}\kern 0.1em{\sc iii}]}}
\newcommand{\MgII}{\hbox{{\rm Mg}\kern 0.1em{\sc ii}}}
\newcommand{\NeIII}{[\hbox{{\rm Ne}\kern 0.1em{\sc iii}}]}

\newcommand{\OIII}{[\hbox{{\rm O}\kern 0.1em{\sc iii}}]}
\newcommand{\OII}{[\hbox{{\rm O}\kern 0.1em{\sc ii}}]}
\newcommand{\CIV}{\hbox{{\rm C}\kern 0.1em{\sc iv}}}
\newcommand{\CII}{[\hbox{{\rm C}\kern 0.1em{\sc ii}}]}

\defcitealias{NaiduMatthee22}{Naidu \& Matthee et al.}

\title[XLS-z2: (Re)Solving Reionization with Ly$\alpha$]{(Re)Solving Reionization with Ly$\alpha$: How Bright Ly$\alpha$ Emitters account for the $z\approx2-8$ Cosmic Ionizing Background}

\author [J.~Matthee \& R.P.~Naidu et al.]{
\parbox[t]{\textwidth}{
Jorryt Matthee$^{1}$\thanks{E-mail: mattheej@phys.ethz.ch, rohan.naidu@cfa.harvard.edu}\thanks{These authors contributed equally to this work.}\thanks{Zwicky Fellow},
Rohan P. Naidu$^{2}$\footnotemark[1]\footnotemark[2], Gabriele Pezzulli$^{3}$, Max Gronke$^{4,5,6}$, David Sobral$^{7}$, Pascal A. Oesch$^{8,9}$, Matthew Hayes$^{10}$, Dawn Erb$^{11}$, Daniel Schaerer$^{8}$, Ricardo Amor\'in$^{12,13}$, Sandro Tacchella$^{14}$, Ana Paulino-Afonso$^{15}$, Mario Llerena$^{12,13}$, Jo\~ao Calhau$^{16,17}$ and Huub R\"ottgering$^{18}$}
\vspace*{8pt}\\
$^{1}$Department of Physics, ETH Z\"urich, Wolfgang-Pauli-Strasse 27, 8093 Z\"urich, Switzerland\\
$^{2}$Center for Astrophysics $|$ Harvard \& Smithsonian, 60 Garden Street, Cambridge, MA 02138, USA\\
See Appendix \ref{appendix:affiliations} for other authors' affiliations.
}

\pubyear{2021}

\begin{document}
\label{firstpage}
\pagerange{\pageref{firstpage}--\pageref{lastpage}}
\maketitle

\begin{abstract}
The cosmic ionizing emissivity from star-forming galaxies has long been anchored to UV luminosity functions. Here we introduce an emissivity framework based on Ly$\alpha$ emitters (LAEs), which naturally hones in on the subset of galaxies responsible for the ionizing background due to the intimate connection between production and escape of Ly$\alpha$ and LyC photons. Using constraints on the escape fractions of bright LAEs ($L_{\rm{Ly\alpha}}>0.2 L^{*}$) at $z\approx2$ obtained from resolved Ly$\alpha$ profiles, and arguing for their redshift-invariance, we show that: (i) quasars and LAEs together reproduce the relatively flat emissivity at $z\approx2-6$, which is non-trivial given the strong evolution in both the star-formation density and quasar number density at these epochs and (ii) LAEs produce late and rapid reionization between $z\approx6-9$ under plausible assumptions. Within this framework, the $>10\times$ rise in the UV population-averaged \fescs between $z\approx3-7$ naturally arises due to the same phenomena that drive the growing Ly$\alpha$ emitter fraction with redshift. Generally, a LAE dominated emissivity yields a peak in the distribution of the ionizing budget with UV luminosity as reported in latest simulations. Using our adopted parameters ($f_{\rm{esc}}=50\%$, $\xi_{\rm{ion}}=10^{25.9}$ Hz erg$^{-1}$ for half the bright LAEs), a highly ionizing minority of galaxies with $M_{\rm UV}<-17$ accounts for the entire ionizing budget from star-forming galaxies. Rapid flashes of LyC from such rare galaxies produce a ``disco'' ionizing background. We conclude proposing tests to further develop our suggested Ly$\alpha$-anchored formalism.
\end{abstract}




\section{Introduction}
The completion of hydrogen reionization by $z\approx6$ marks the last major phase transition of the Universe \citep[e.g.][]{Fan06, Loeb13, McGreer15, McQuinn16, Stark16, Wise19, Dayal20}. Due to the low number density of quasars, the cosmic ionizing emissivity was likely dominated by star-forming galaxies at $z>6$ \citep[e.g.,][]{Matsuoka18, Parsa18, Kulkarni19, Shen20}. While this broad picture is in place, major open questions remain. Was reionization driven by multitudes of ultra-faint ($L_{\rm{UV}}<0.01 L^{*}$) galaxies below current observational limits, or by relatively rare, bright ($L_{\rm{UV}}>0.1 L^{*}$) galaxies that are already being catalogued \citep[e.g.,][]{Finkelstein19,Naidu20}? An intertwined puzzle is why after reionization is complete, the ionizing background seems to remain approximately flat across $z\approx2-6$ \citep[e.g.,][]{Becker13,Khaire19,FaucherGiguere20} -- this is in sharp contrast with both the rapidly changing star formation rate density \citep[e.g.,][]{Bouwens21} as well as quasar number density \citep[e.g.,][]{Kulkarni19} during this epoch. Another related conundrum is the large Ly$\alpha$ opacity fluctuations observed at $z\approx5-6$, which have been interpreted as evidence for extended and patchy reionization.  The precise origin of these fluctuations are as yet unknown \citep[e.g.,][]{Keating20,Bosman21,Zhu21}. All these issues may be summarized as follows: we are in need of a unified, empirically grounded origin story for the $z\approx2-8$ cosmic ionizing background.

The first place to start is by re-examining our tools, and the implicit assumptions steeped in them. Calculations of the ionizing emissivity from star-forming galaxies typically adopt the following UV luminosity anchored formalism \citep[e.g.,][]{Madau99,Robertson13,duncan&conselice15}: 

\begin{equation}
\label{eq:nion}
    \dot{n}_{\mathrm{ion}}(z) = \rho_{\mathrm{UV}}(z)\ \xi_{\mathrm{ion}}\ f^{\rm{LyC}}_{\mathrm{esc}}.
\end{equation}

The ionizing emissivity $\dot{n}_{\mathrm{ion}}(z)$ is intuitively expressed as the product of the UV luminosity density ($\rho_{\mathrm{UV}}$; \citealt[][]{Bouwens15aLF,Finkelstein16,Oesch18}), a global UV luminosity to ionizing photon conversion factor ($\xi_{\mathrm{ion}}$; \citealt{Matthee17,Shivaei18,Lam19}), and the ionizing photon escape fraction (\fesc\footnote{The escape fraction that enters this equation is the `globally averaged' escape fraction \citep[see e.g.][]{HaardtMadau2012}, which averages over the number densities and ionising luminosity of the galaxy population. Throughout the paper we refer to population-averaged escape fractions as $\langle f_{\rm esc}\rangle$ in case this is particularly emphasised.}; e.g. \citealt{Marchi17,Naidu18,Steidel18}).
The preeminence of this formalism is because UV luminosity functions (LFs) represent the most complete, homogeneously gathered samples of star-forming galaxies known at $z=3-10$.

However, the UV luminosity is only tenuously linked to escaping ionizing photon luminosity. This is because the UV luminosity is related to variations in star-formation on 100 Myr timescales \citep[e.g.][]{Caplar19}, while ionizing photon escape is likely a stochastic, bursty, feedback-driven process varying on shorter timescales (\citealt{Rosdahl18,Kimm19,Ma20}). Furthermore, the UV luminosity is only mildly sensitive to the processes that plausibly influence \fescs (such as the HI column density and dust; e.g. \citealt{Chisholm18}). 

In fact, the majority of UV-selected galaxies may contribute very little to the ionizing background at $z<6$. This follows directly from these premises: (i) the fraction of Ly$\alpha$ photons that escape is strictly greater than or equal to the Lyman Continuum (LyC) \fescs \citep[e.g.][]{Dijkstra16,Izotov2020} so galaxies without escaping Ly$\alpha$ emission likely have LyC \fesc$\approx0$, and (ii) the majority of galaxies are not Ly$\alpha$ emitters (LAEs; at least out to $z\approx6$ and $M_{\rm{UV}}<-19$; \citealt{Stark16, Kusakabe2020,Ouchi2020,Santos21}). However, in the canonical UV-anchored formalism these details are all buried in the population-averaged escape fraction and its redshift dependence, i.e. $\langle f_{\rm{esc}}\rangle(z)$, which is extremely challenging to measure at high-redshift due to IGM absorption and foreground projections \citep[e.g.][]{Vanzella10,Siana15,Jones18}. In order to reconcile the evolution of $\dot{n}_{\mathrm{ion}}(z)$ with measurements of the emissivity from quasar absorption line statistics at $z<6$, neutral fractions at $z>6$, and the \citet{Planck18} Thomson optical depth, one needs to assume strong redshift evolution in the average \fesc, approximately $\propto(1+z)^{3}$ \citep[e.g.,][]{HaardtMadau2012,Robertson15,Puchwein19,Mason19,FaucherGiguere20}. Explaining the origin of this strongly evolving escape fraction is a key challenge.

In this paper we aim to explain the relative flatness of the ionizing background by exploring a galaxy emissivity dominated by LAEs. The driving motivation is that both empirical and theoretical work reveal the escape of Ly$\alpha$ emission to be intimately linked with LyC \fescs \citep[e.g.][]{Steidel18,Izotov18b,Marchi18, Verhamme15,Dijkstra16,Gronke17}. Thus, LAEs naturally represent the subset of star-forming galaxies that actively contribute to the ionizing background. Indeed, practically all known galaxies with significant (e.g. $>5$\%) LyC leakage are strong Ly$\alpha$ emitters (e.g. \citealt{Rivera-Thorsen17,Vanzella18,Izotov21}, but see the discussion in \citealt{Ji20} for one possible counter example)\footnote{There is so far no evidence that extreme leakers that emit negligible nebular recombination lines exist (for a more detailed discussion we refer to \S7 of \citetalias{NaiduMatthee22} \citeyear{NaiduMatthee22}).}. This implies that the shape of the (escaping) LyC LF is much more tightly connected to the shape of the Ly$\alpha$ LF \citep[e.g.][]{Sobral18,Konno18,Herenz19}, than it is to the shape of the UV LF.

At $z\approx2-3$, where detailed rest-frame optical spectroscopy is possible, studies have shown that LAEs are typically relatively young galaxies with low dust content, high ionisation states and high specific SFRs \citep{Nakajima16,Trainor16,Matthee21,Reddy21}. These properties are expected to be common among galaxies in the Epoch of Reionisation (EoR) and indeed, the fraction of UV-selected Lyman Break Galaxies (LBGs) that are LAEs increases steadily with redshift \citep[e.g.][]{Stark11,Kusakabe2020,Ouchi2020}. This is also reflected by the relative flatness of the Ly$\alpha$ luminosity density during $z=3-6$ \citep[e.g.,][]{Dawson07,Ouchi08,Sobral18,Herenz19}, an epoch over which the UV luminosity density significantly increases \citep[][e.g.]{Hayes2011,Sobral18}. The falling cosmic star-formation rate density \citep[e.g.,][]{madau&dickinson} is counterbalanced by the increasing incidence of Ly$\alpha$ emitters towards higher redshift \citep[e.g.,][]{Stark11}, by about a factor ten. A consequence is that the average Ly$\alpha$ escape fraction of the \textit{full} galaxy population at $z\gtrsim6$ is comparable to the average Ly$\alpha$ escape fraction of LAEs at $z=2$ \citep{Matthee21}. The processes that determine the Ly$\alpha$ escape in these LAEs at $z=2$ are likely present in the average galaxy in the EoR.

Resolved Ly$\alpha$ line profiles allow us to select the specific galaxies that are in the LyC leaking phase \citep[e.g.][]{Verhamme15,Dijkstra16}, and to infer their \fescs \citep[e.g.][]{Izotov18b,Gazagnes20}. In a companion paper (\citetalias{NaiduMatthee22} \citeyear{NaiduMatthee22}) we showed that half of the bright LAEs selected from the X-SHOOTER Lyman-$\alpha$ survey at $z\approx2$ (XLS-$z2$; \citealt{Matthee21}; LAEs with a luminosity $>0.2$ L$^{\star}$) have Ly$\alpha$ line-profiles suggestive of being in a phase where they are leaking ionising photons with an escape fraction in the range 20-50 \%. The other half of the LAEs have Ly$\alpha$ photons that only escape after scattering to the edge of the line profile due to a high HI column density and they therefore have negligible \fescs of ionising photons. The galaxies that have LyC leakage are also the ones that simultaneously have a very high ionising photon production efficiency. Here we place these measurements in a statistical framework and infer the cosmic emissivity due to LAEs at $z\approx2$ and extrapolate these findings out to $z=8$.

In \S $\ref{sec:formalism}$ we present the derivation of a Ly$\alpha$-based emissivity formalism. We then discuss the knowns and unknowns of the parameters that are invoked in the formalism to compute the emissivity from LAEs over $z=2-8$ in \S $\ref{sec:knowns}$. In \S $\ref{sec:knowns}$ we also motivate the choices of our simple fiducial model parameters such as the \fescs and the luminosity limit of the faintest contributor to the emissivity. The main results based on this fiducial model are demonstrated in \S $\ref{sec:results}$, which first shows the total ionizing emissivity at $z\approx2-8$ due to LAEs and quasars and then focuses on the role of LAEs in reionising the Universe at $z>6$. We connect the results from the LAE-formalism to the general galaxy population in \S $\ref{sec:LBGframework}$. We discuss the implications of our results in \S\ref{sec:discussion}, addressing the open questions raised at the beginning of this section. In Appendix $\ref{sec:modelvar}$ we show which results and aspects of the LAE-emissivity model are sensitive to choices in the key parameters.

Throughout this work we reference $L^{*}$, the characteristic luminosity in Schechter function parametrizations of luminosity functions (LFs). In the context of Ly$\alpha$ LFs, $L^{*}$ is as per the \citet{Sobral18} $z\approx2-6$ consensus LFs ($\log{L_{\rm{Ly\alpha}}/\rm{erg\ s^{-1}}}\approx43$) and in the context of LBGs it is M$^{\star}_{\rm UV}\approx-21$ as per the \citet{Bouwens21} LFs. We assume a flat $\Lambda$CDM concordance cosmology with $\Omega_M=0.3$, $\Omega_{\Lambda}=0.7$ and $H_0=70$ km s$^{-1}$ Mpc$^{-1}$.

\begin{figure*}
\centering
\begin{tabular}{cc}
\includegraphics[width=0.5\linewidth]{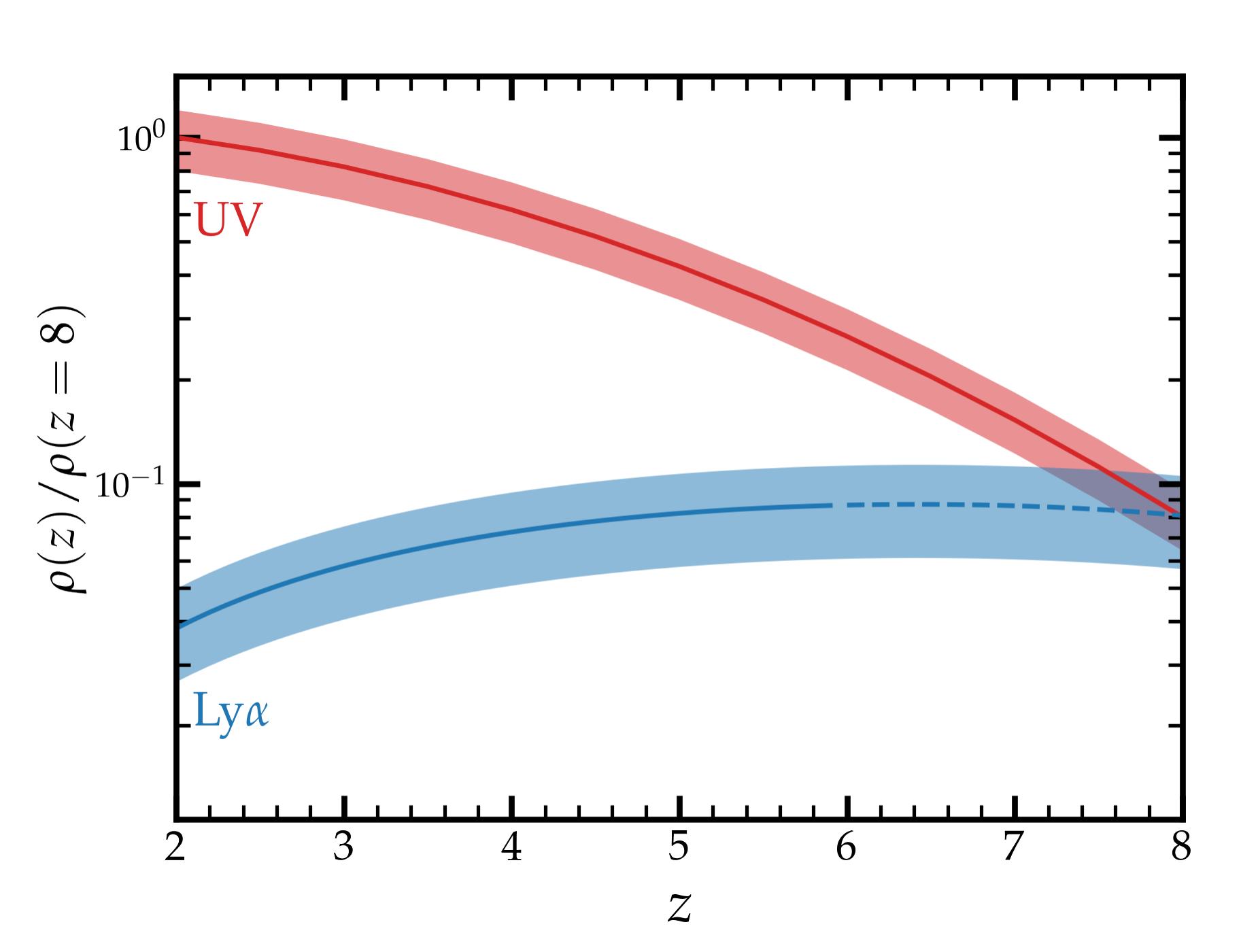} &
\hspace{-0.5cm}\includegraphics[width=0.5\linewidth]{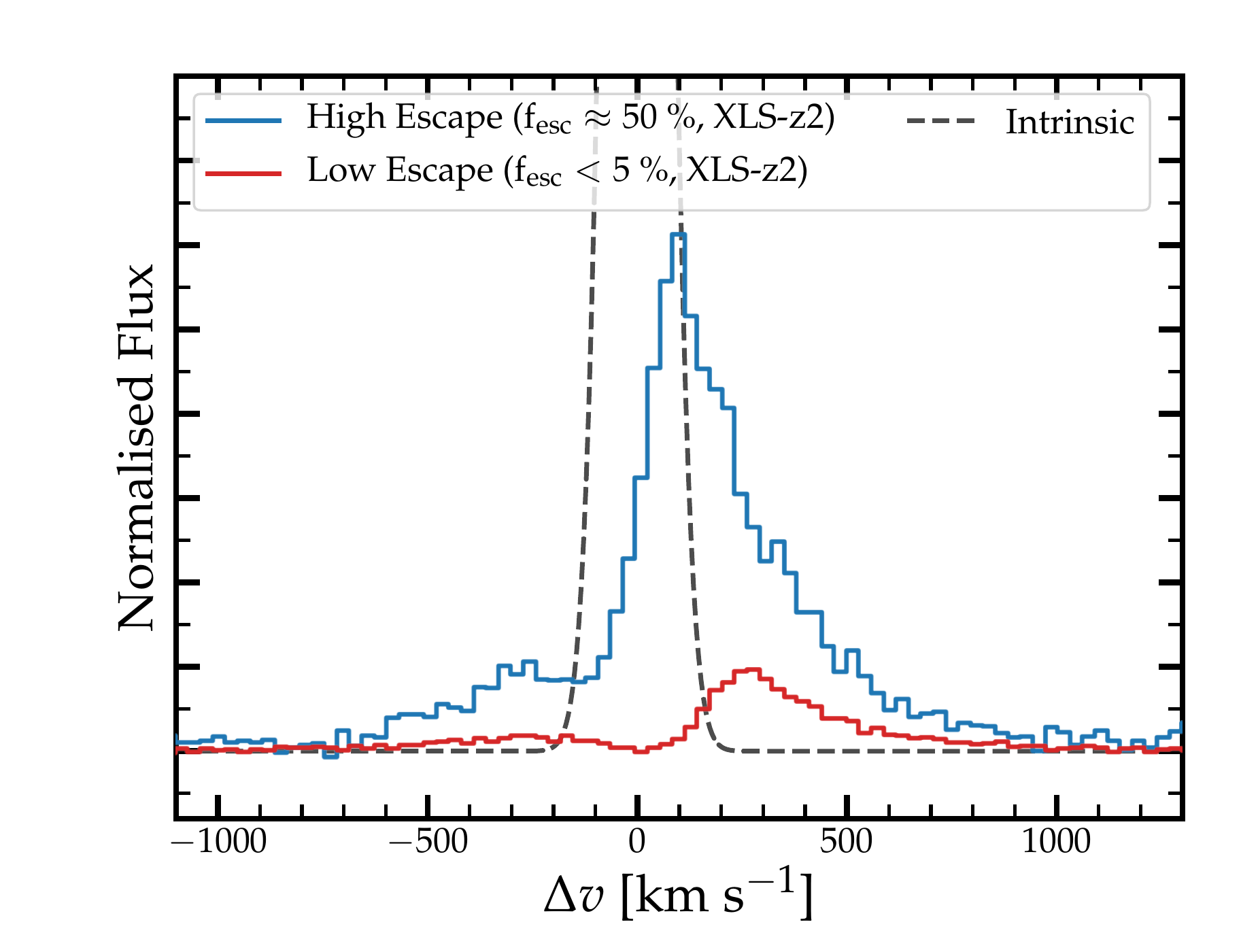}
\end{tabular}
\caption{\textbf{Left:} The relative flatness of the integrated Ly$\alpha$ luminosity density (blue; interpolating Ly$\alpha$ luminosity functions listed in Table $\ref{tab:knowns}$ integrated down to a luminosity $10^{42.2}$ erg s$^{-1}$) is in stark contrast to the strong decline in the UV luminosity density with redshift (red; from \citealt{Bouwens21}, integrated down to M$_{\rm UV}=-17$). Shaded regions show the propagated uncertainties. We extrapolate the evolution of the Ly$\alpha$ luminosity density above $z>6$ due to the possible significant impact of the IGM on published measurements at $z>6$ (see $\ref{sec:reionization}$). The luminosity densities are scaled to their densities at $z=8$. The difference between the Ly$\alpha$ and the UV luminosity density highlights the evolution in the cosmic averaged Ly$\alpha$ escape fraction (e.g. \citealt{Hayes2011}). \textbf{Right:} The Ly$\alpha$ profiles of LAEs at $z\approx2$ for LAEs with high (blue) and low (red) \fescs (see \citetalias{NaiduMatthee22} \citeyear{NaiduMatthee22}), respectively. The line-profiles are normalised to the intrinsic Ly$\alpha$ emission estimated from their specific dust-corrected H$\alpha$ luminosity. For leaking LAEs, a higher fraction of Ly$\alpha$ flux escapes and photons do this much nearer to line-center.}
\label{fig:intro}
\end{figure*}

\section{A \lyas based emissivity formalism} \label{sec:formalism}
Here we derive the emissivity from LyC leaking LAEs, i.e. the average HI ionizing luminosity density that escapes from LAEs per unit comoving volume. The emissivity at a given redshift depends on the number density of LAEs and their ionizing output. In analogy to the UV-continuum approach \citep[e.g.][]{Robertson13}, we formulate the emissivity due to LAEs, $\dot{n}_{\mathrm{ion, LAE}}(z)$, as follows:

\begin{equation}
\label{eq:nionlya}
    \dot{n}_{\mathrm{ion, LAE}}(z) = \rho_{\mathrm{Ly \alpha}}(z)\ \xi_{\mathrm{ion}}^{\rm{Ly\alpha}}\ f^{\rm{LyC}}_{\mathrm{esc}}.
\end{equation}

The \lyas luminosity density, $\rho_{\mathrm{Ly \alpha}}(z)$, replaces $\rho_{\mathrm{UV}}(z)$ from Eqn. \ref{eq:nion}. $f_{\rm{esc}}^{\rm{LyC}}$ represents the LyC escape fraction of LAEs. This leaves us with $\xi_{\mathrm{ion}}^{\rm{Ly\alpha}}$, which converts the observed Ly$\alpha$ luminosity into a number of ionizing photons produced per second:
\begin{equation} \label{eq:xi}
    \xi_{\mathrm{ion}}^{\rm{Ly\alpha}} = \frac{Q_{\rm{ion}}}{L_{\rm{Ly\alpha, obs}}},
\end{equation}
where $Q_{\rm{ion}}$ expresses the total number of ionizing photons being produced per second. 
The observed Ly$\alpha$ luminosity can be related to the intrinsic H$\alpha$ luminosity produced in an HII region, and thus $Q_{\rm ion}$ if we know the Ly$\alpha$ escape fraction, $f_{\rm{esc}}^{\rm{Ly\alpha}}$ \citep[e.g.][]{Sobral19}. Here, we ignore collisionally excited emission. Under standard Case B recombination assumptions ($T=10^{4}$ K, $n_{e}=350$ cm$^{-3}$), $L_{\rm{Ly\alpha, int}}/L_{\rm{H\alpha, int}} = 8.7 $, with typical values for Green Peas and LAEs expected to range from $\approx8-9$ \citep[e.g.,][]{Henry15}. Therefore:

\begin{align}
\label{eq:lyafesc}
f_{\rm{esc}}^{\rm{Ly\alpha}} & = \frac{L_{\rm{Ly\alpha, obs}}}{L_{\rm{Ly\alpha, int}}}  = \frac{L_{\rm{Ly\alpha, obs}}}{8.7 L_{\rm{H\alpha, int}}},
\end{align}

Having obtained the intrinsic H$\alpha$ luminosity from measurements of $f_{\rm{esc}}^{\rm{Ly\alpha}}$, we can calculate the number of produced ionizing photons as follows:

\begin{equation}
\label{eq:lha1}
L_{\rm{H\alpha, int}} = Q_{\rm{ion}} \left(1 - f_{\rm{esc}}^{\rm{LyC}} - f_{\rm{dust}}\right) c_{\rm{H\alpha}},
\end{equation}
where we note that some fraction of the produced ionizing photons escape the galaxy ($f_{\rm{esc}}^{\rm{LyC}}$) and some fraction are absorbed by dust ($f_{\rm{dust}}$). The remaining fraction of photons ($1-f_{\rm{esc}}^{\rm{LyC}}-f_{\rm{dust}}$) is available to produce nebular emission. Since we are interested in the intrinsic luminosity of H$\alpha$ ($L_{\rm{H\alpha,int}}$), i.e., the luminosity before attenuation, we multiply this remaining fraction of photons with the H$\alpha$ line emission coefficient $c_{\rm{H\alpha}}$ whose value is $\approx1.25-1.35\times10^{-12}$ erg for typical Case B recombination assumptions ($T=10^{4}$ K, $n_{e}=350$ cm$^{-3}$, \citealt[][]{Kennicutt98,Schaerer03}). For the galaxies of interest (galaxies with f$_{\rm esc}>0$), we assume $f_{\rm{dust}}\approx0$ given that their nebular attenuation seems to be negligible (\citetalias{NaiduMatthee22} \citeyear{NaiduMatthee22}). We refer to \cite{Kakiichi19} for a more detailed discussion on the role of dust within HII regions in the context of Ly$\alpha$ and LyC escape.

We can then express $\xi_{\mathrm{ion}}^{\rm{Ly\alpha}}$ in terms of these various escape fractions: 
\begin{align}
\label{eq:xiionlya}
\xi_{\mathrm{ion}}^{\rm{Ly\alpha}} & = \frac{Q_{\rm{ion}}}{L_{\rm{Ly\alpha, obs}}}\nonumber\\
& = \frac{1}{8.7 f_{\rm{esc}}^{\rm{Ly\alpha}}\left(1 - f_{\rm{esc}}^{\rm{LyC}} - f_{\rm{dust}}\right)c_{\rm{H\alpha}}}.
\end{align}
Combining Eqn. $\ref{eq:xiionlya}$ with Eqn. $\ref{eq:nionlya}$ finally yields:

\begin{align}
\label{eq:nionlyasummary}
    \dot{n}_{\mathrm{ion, LAE}}(z) &=  \frac{\rho_{\mathrm{Ly \alpha}}(z) f^{\rm{LyC}}_{\mathrm{esc}}}{8.7 f_{\rm{esc}}^{\rm{Ly\alpha}}\left(1 - f_{\rm{esc}}^{\rm{LyC}}\right)c_{\rm{H\alpha}}}.
\end{align}

In our implementation of the formalism based on our observational results described in \citetalias{NaiduMatthee22} (\citeyear{NaiduMatthee22}), we split the sample of LAEs in those that have LyC leakage and those that have not. This means that we specifically calculate the emissivity as:

\begin{align}
\label{eq:nionlyasummary2}
    \dot{n}_{\mathrm{ion, LAE}}(z) &=  \frac{\rho_{\mathrm{Ly \alpha}}(z) \, f_{\rm{LyC, LAEs}} \, f^{\rm{LyC}}_{\mathrm{esc}}}{8.7 f_{\rm{esc}}^{\rm{Ly\alpha}}\left(1 - f_{\rm{esc}}^{\rm{LyC}}\right)c_{\rm{H\alpha}}},
\end{align}
where $f_{\rm{LyC, LAEs}}$ is the fraction of LyC leaking LAEs with the respective LyC and Ly$\alpha$ escape fractions $f_{\rm{esc}}^{\rm{LyC}}$ and $f_{\rm{esc}}^{\rm{Ly\alpha}}$. Because of the non-linear correction $\propto\frac{1}{1-f_{\rm{esc}}^{\rm{LyC}}}$, we note that this implementation is not exactly the same as when we would use the population-averaged Ly$\alpha$ and LyC escape fractions for all LAEs in Eq. $\ref{eq:nionlyasummary}$, but the differences are very minor (within 0.05 dex). 

For comparison with $z<6$ constraints it is convenient to work with $\epsilon_{\rm{912}}$ -- the emissivity at 912$\AA$ -- which is more directly inferred from the Ly$\alpha$ forest \citep[e.g.,][Eqn. 7]{Becker13}.

\begin{align}
\label{eq:e912}
    \epsilon_{\mathrm{912}}(z) &= \dot{n}_{\mathrm{ion, LAE}}(z) h \alpha,
\end{align}
where $h$ is Planck's constant, and $\alpha$ is the spectral slope such that the total number of ionizing photons is given by the integral $\int_{\rm{\nu_{\rm{912}}}}^{\infty} \frac{d\nu}{h\nu}\epsilon_{\mathrm{912}}\left(\frac{\nu}{\nu_{\rm{912}}}\right)^{-\alpha}$. For the spectral template adopted in \citetalias{NaiduMatthee22} (\citeyear{NaiduMatthee22}) to describe the stellar ionizing spectrum of LyC leakers, $\alpha=1.25$. The Ly$\alpha$ forest studies typically convert their inferred $\epsilon_{\rm{912}}$ to an $\dot{n}_{\rm{ion}}$ by assuming some $\alpha$, but the appropriate $\alpha$ that must be adopted is unclear at $z<6$ where both AGN and LAEs contribute, and their relative contributions as well as spectral shapes are uncertain. We note that we do not explicitly investigate the impact of LAEs on the HeII-ionizing emissivity because the HeII-ionizing luminosity from star-forming galaxies is small, even for the strongly ionizing stellar spectra that power the LAEs.

\section{Computing the LAE emissivity: knowns and unknowns} \label{sec:knowns}

In this section we discuss the knowns and unknowns in the Ly$\alpha$-anchored emissivity framework (i.e., Eqn. \ref{eq:nionlyasummary}) for LAEs over $z=2-8$ (summarised in Table $\ref{tab:knowns}$). Throughout this paper we use a simple fiducial calculation that demonstrates the applicability of the LAE-framework. As a fiducial choice of parameters we explicitly explore the contribution of the currently observed population of bright LAEs ($L_{\rm Ly\alpha}>10^{42.2}$ erg s$^{-1}$, $\approx0.2$ L$^{\star}$, the luminosity limit of our XLS-$z2$ survey; \citealt{Matthee21}) for which the Ly$\alpha$ LF is determined over $z\approx2-7$ and for which we have constrained the fraction of LyC leakers and their escape fraction in \citetalias{NaiduMatthee22} (\citeyear{NaiduMatthee22}). We assume that the contribution of fainter LAEs to $\dot{n}_{\mathrm{ion, LAE}}(z)$ is negligible (i.e. their \fescs=0), which is equivalent to introducing a cut-off in the luminosity function.  We present the implications of a scenario in which the cut-off luminosity is a factor ten lower in Appendix $\ref{sec:modelvar}$ and include this in our Discussion. Our aim is to explore whether we can explain the evolution of the neutral fraction in the EoR, and the following evolution of the cosmic ionising background with these bright LAEs when combined with the quasar contribution \citep{Kulkarni19}. The relevant parameters chosen in our fiducial model are empirically constrained and summarised in Table $\ref{tab:knowns}$. The results of this calculation are presented in \S $\ref{sec:results}$. 

We emphasize that our fiducial calculation does not extrapolate number densities or ionizing properties beyond the luminosity range for which these ionizing properties have been constrained at $z\approx2$ (i.e., $L_{\rm{Ly\alpha}}>0.2 L^{*}$, i.e. the selection function of our spectroscopic follow-up survey of LAEs). Contrast this with the UV-anchored approach where \fescs and $\xi_{\rm{ion}}$ are extrapolated $\approx100\times$ fainter than the faintest galaxies for which these quantities have been inferred (typified by Model I in \citealt{Naidu20}).

\subsection{Ly$\alpha$ luminosity density} \label{sec:LyaLF}
Ly$\alpha$ luminosity functions are constrained from $\approx0.1-10 L^{*}$ over various redshift slices from $z\approx2-7$  \citep[e.g.,][]{Ouchi18,Sobral18,Hu19,delavieuville19,Taylor20}. In our estimate, we compute $\rho_{\rm{Ly\alpha}}$ by integrating LFs based on large narrow-band surveys \citep{Ouchi2020}, recent data from integral field spectroscopy \citep[IFU; e.g.][]{Herenz17}, and the ``S-SC4K" consensus LFs \citep{Sobral18} that combine IFU and narrow-band data. The benefit of the narrow-band surveys is that they probe larger volumes and narrow specific redshift slices, while the benefit of IFU surveys \citep[e.g.][]{Bina16,Drake17b} is that they have full spectroscopically confirmed samples and probe fainter luminosities. As can be seen from the LAE emissivities listed in Table $\ref{table:emissivity}$, the integrated Ly$\alpha$ luminosity densities of various published LFs at similar redshifts agree within the uncertainties despite that their shapes are somewhat different \citep[c.f.][]{Santos16,Konno18}. Note that a very minor AGN contribution to $\rho_{\rm{Ly\alpha}}$ is not excluded at $z\gtrsim5$, as AGN can plausibly not be fully removed from the Ly$\alpha$ LFs as the currently available X-Ray data is not sensitive enough \citep[e.g.][]{Calhau20}. The AGN fractions among LAEs at $z\approx2-3$ are however only significant for $\gg L^{\star}$ luminosities \citep{Matthee17Bootes,Calhau20}, and appear to decrease at $z>3$ at fixed Ly$\alpha$ luminosity \citep[e.g. Fig. 9 in][]{Sobral18} suggesting that the AGN contribution to $\rho_{\rm{Ly\alpha}}$ is low at $z\gtrsim5$ as well. This is discussed further in connection to the shape of the Ly$\alpha$ LF in Appendix $\ref{sec:AGN}$.

An important caveat for the Ly$\alpha$ framework is the impact of IGM transmission on the observed Ly$\alpha$ line luminosity. The increasing density of neutral absorbers with redshift likely reduces the observed Ly$\alpha$ transmission from galaxies \citep[e.g.][]{Laursen2011,Smith21}. . The magnitude of the IGM transmission is uncertain as the transmission depends on large scale motions of gas around galaxies, the neutral fraction of the IGM and the proximity zones around galaxies \citep[e.g.][]{Gronke20,Park21}. The presence of large opacity fluctuations in quasar spectra at $z\sim6$ suggests that significant neutral patches may have survived down to even $z\approx5.5$ \citep[e.g.][]{Kulkarni19b,Keating20}, impacting Ly$\alpha$ observability from galaxies \citep{Christenson21}. Moreover, the effective IGM correction is strongly sensitive to the Ly$\alpha$ line-profile as it escapes the galaxy \citep[typically peaking at a velocity offset of $\approx+200$ km s$^{-1}$ at $z>3$, e.g.][]{Cassata2020}, and the effective correction that should be applied is the one applicable to `favourable sight-lines' due to the Ly$\alpha$ pre-selection of the sources that are used to measure the luminosity function.

Considering these complexities, in our fiducial model, we adopt empirical transmission corrections based on the results from \cite{Hayes21}, who show that the evolution in the blue parts of average Ly$\alpha$ line-profiles can be explained with corrections assuming average IGM sight lines \citep{Inoue14}, without requiring corrections for the red part of the line. The corrections for the total Ly$\alpha$ luminosities are a factor 1.0 at $z=2$ to $1.2$ at $z=6$ and we assume that they are the same for all LAEs at a fixed redshift. At $z>6$, when multiple probes agree reionization has not yet been completed, these IGM corrections are likely too low partially due to absorption of the damping wing in the presence of extended neutral regions, and even the red side of Ly$\alpha$ can be damped \citep[e.g.][]{Jung20}. Therefore we are able to report only lower limits on the emissivity, as our IGM corrections may significantly under-estimate the Ly$\alpha$ luminosity density emitted from galaxies at these redshifts. We make a finer extrapolation of the IGM-corrected emissivity into the EoR in \S\ref{sec:reionization}.

After applying these small IGM corrections, we integrate the Ly$\alpha$ LFs (parametrised by Schechter functions) down to a luminosity limit of $L_{\rm Ly\alpha}>10^{42.2}$ erg s$^{-1}$ ($\approx 0.2$ L$^{\star}$; \citealt{Sobral18}). The evolution of the integrated luminosity density down to this limit is shown in the left panel of Fig. $\ref{fig:intro}$. Unlike the UV luminosity density, there is little evolution in the Ly$\alpha$ luminosity density over the $z\approx3-6$ interval.

\begin{table*}
    \centering
    \caption{Summary of the key input parameters for Eqn. $\ref{eq:nionlyasummary}$ used in our {\it fiducial} calculation of the emissivity from LAEs. }

    \begin{tabular}{l|p{9.5cm}} \hline
        {\bf Ly$\alpha$ luminosity density:} & $\rho_{\mathrm{Ly \alpha}}(z)$  \\
        Ly$\alpha$ LF & Well-determined for $>0.1$ L$^{\star}$ over $z=2-6$ \citep{Sobral18,Herenz19,Ouchi2020}, but possibly affected by neutral IGM at $z>6$ \citep{Santos16,Konno18,Wold21}. \\
      IGM correction & We assume empirical corrections that are shown to yield consistent Ly$\alpha$ line profiles over $z=3-6$ by \citet{Hayes21}. Corrections are particularly uncertain at $z\gtrsim6$.  \\
        Integration limit & We integrate LFs down to a limiting luminosity of L$_{\rm Ly\alpha}>10^{42.2}$ erg s$^{-1}$ which is the limit of the XLS-$z2$ sample for which we determined $\langle f_{\rm esc} \rangle$. We assume the contribution of fainter LAEs to $\dot{n}_{\mathrm{ion, LAE}}(z)$ to be negligible (i.e. their \fescs=0). \\ \hline
        
        {\bf Escape Fractions:} & $f_{\rm esc}= 50$ \% for half the LAEs with L$_{\rm Ly\alpha}>10^{42.2}$ erg s$^{-1}$, independent of redshift \\
        Fraction of LyC-leaking LAEs, $f_{\rm{LyC, LAEs}}$ & Estimated to be $50\pm10$ \% based on Ly$\alpha$ line-profile statistics (\citetalias{NaiduMatthee22} \citeyear{NaiduMatthee22}). \\
        $f_{\rm esc}$ for LyC-leaking LAEs & Assumed to be 50 \% for all leaking LAEs, the fiducial \fescs for this sample as discussed in \citetalias{NaiduMatthee22} (\citeyear{NaiduMatthee22}). \\
        Redshift evolution & These two quantities are assumed to be redshift invariant based on the observed invariance of average Ly$\alpha$ line-profiles \citep{Hayes21}. We assume luminosity independence. \\
        $f_{\rm esc, Ly\alpha}$ for LyC-leaking LAEs & Assumed to be $\approx50\%$ based on measurement of $47^{+3}_{-8}$ \% for LyC-leaking LAEs (\citetalias{NaiduMatthee22} \citeyear{NaiduMatthee22}).  \\\hline

        {\bf Extrapolation into the EoR at $z>6$} (\S $\ref{sec:reionization}$) &  \\
        Evolution of IGM-unaffected Ly$\alpha$ LF & At $z>6$, $\dot{n}_{\rm{ion}}(z) = \dot{n}_{\rm{ion}}(z=5.7) \times  \frac{\Phi^{\star}_{\rm LBG}(z) X_{\rm{LAE}}(z)}{\Phi^{\star}_{\rm LBG}(z=5.7) X_{\rm{LAE}}(z=5.7)}$  where $\Phi^{\star}_{\rm LBG}$ corresponds to the \citet{Bouwens21} UVLFs and $X_{\rm{LAE}}$ is the fraction of LBGs with L$_{\rm Ly\alpha}>10^{42.2}$ erg s$^{-1}$ extrapolated from the observed trend at $z=3-6$ \citep{Santos21}, and capped to $X_{\rm{LAE}}=1$. \\ \hline
    \end{tabular}
    \label{tab:knowns}
\end{table*}

\subsection{Escape fractions} \label{sec:fesc}
The escape fraction of ionising photons in LyC leaking LAEs is the most uncertain quantity determining $\dot{n}_{\mathrm{ion, LAE}}(z)$, and it is also very important because of its $\propto \frac{f_{\rm esc}}{1-f_{\rm esc}}$ dependence. Empirically, it is observed that the LyC \fescs is very similar to the \lyas \fescs in LyC leakers with a high escape fraction \citep[e.g.][]{Izotov2020}. Recently, the unique combination of high resolution Ly$\alpha$ spectra with available systemic redshifts and the well-characterised selection function of the XLS-$z2$ survey allowed us to use the Ly$\alpha$ line profile to identify which galaxies are strong leakers and to indirectly infer the escape fraction of these LAEs. Half the LAEs ($f_{\rm{LyC, LAEs}}=50\pm10$ \%) at $z\approx2$ show Ly$\alpha$ profiles that are comparable to confirmed LyC leakers with $f_{\rm esc}>$20 \% (\citetalias{NaiduMatthee22} \citeyear{NaiduMatthee22}). The Ly$\alpha$ photons in these LAEs escape particularly close to the systemic redshift, and they have narrow peak separations (right panel of Fig. $\ref{fig:intro}$). The average \fescs of this half of LAEs is estimated to be in the range $20-50$ \% from the Ly$\alpha$ line profiles and the average Ly$\alpha$ escape fraction of the LyC leakers is measured from the dust-corrected H$\alpha$ to Ly$\alpha$ ratio to be $47^{+3}_{-8}$ \%. This value is used in the fiducial model. The other LAEs in the XLS-$z2$ sample very likely has a negligible \fescs as their wide Ly$\alpha$ profiles shown in Fig. $\ref{fig:intro}$ imply relatively large column densities of neutral hydrogen (see \citetalias{NaiduMatthee22} \citeyear{NaiduMatthee22} for details). The Ly$\alpha$ escape fraction for the non-leaking LAEs is $9^{+2}_{-2}$ \% which implies that the average Ly$\alpha$ escape of all LAEs in the parent sample is $\approx30$ \%, consistent with \citet{Trainor16,Sobral2017,Matthee21}. 

For our fiducial model, we adopt LyC \fesc$\ =\ $\lyas \fesc$=50\%$ for half the LAEs with $L_{\rm Ly\alpha}>10^{42.2}$ erg s$^{-1}$, and zero for all other LAEs. The fiducial choice of this high escape fraction is motivated in detail in Section 6 of \citetalias{NaiduMatthee22} (\citeyear{NaiduMatthee22}) based on three arguments. First, in known LyC leakers at low-redshift, the LyC \fescs approaches the Ly$\alpha$ \fesc, in particular for leakers with a comparably narrow Ly$\alpha$ profile as the leakers in the XLS-$z2$ sample. Second, the shape of the Ly$\alpha$ line profile of the leakers suggests such a high escape fraction when comparing to the known LyC leakers, in particular due to significant fraction of Ly$\alpha$ photons escaping at the systemic redshift \citep[e.g.][]{Gazagnes20}. Thirdly, the fiducial LyC \fescs implies a population-averaged $\langle f_{\rm esc}\rangle\approx25\pm5$ \% for LAEs, which is comparable to the averaged \fescs measured in stacks of LAEs at $z=3$ with similar brightness \citep{Pahl21}, who report $25\pm3$ \% (see Appendix $\ref{sec:Pahl}$ for more comparisons to the results from \citealt{Pahl21} including the population averaged \fescs of the full galaxy population).\footnote{We note that the escape of Ly$\alpha$ and LyC photons is likely anisotropic, and therefore the escape fractions along the line of sight are impacted by the viewing angle \citep[e.g.,][]{Gnedin08b,Wise&Cen09,Paardekooper15,Smith2019}. In our framework, stochastic viewing angle effects that impact the Ly$\alpha$ escape fraction along the line of sight are naturally accounted for in the difference between the Ly$\alpha$ and UV LFs (i.e. the Ly$\alpha$ emitter fraction). However, it is unclear whether Ly$\alpha$ and LyC escape are subject to the same systematics. In case the Ly$\alpha$ line-profile is not only sensitive to \fescs along our line of sight, but also to escaping ionizing photons in another direction, we could possibly over-estimate the population (or viewing-angle)-averaged \fescs using Ly$\alpha$ line-profile statistics. The similar $\langle f_{\rm esc}\rangle\approx25\pm5$ \% implied by our Ly$\alpha$ line-profil studys and direct LyC stacking in LAEs \citep{Pahl21} suggests that such {\it differential} viewing angle effects are not very strong.} We discuss the consequence of including a possible contribution of fainter LAEs combined with a lower $\langle f_{\rm esc} \rangle$ for LAEs in \S $\ref{sec:lowerlimit}$.

\subsection{Redshift invariance of the ionizing properties of LAEs}
In order to calculate the LAE-emissivity at $z>2$  we assume that our estimated Ly$\alpha$ escape fraction, the LyC escape fraction and the fraction of LyC leakers among LAEs at $z\approx2$ are redshift-invariant for LAE populations. The invariance of $f_{\rm esc,LyC}$ and $f_{\rm{LyC, LAEs}}$ is motivated by the lack of evolution in the average Ly$\alpha$ profile of $\gtrapprox0.1 L^{*}$ LAEs over $z\approx2-6$ \citep{Hayes21} and the strong connection between the shape of the Ly$\alpha$ line and \fescs \citep{Verhamme15,Izotov18b}. Furthermore, observations show that several fundamental properties of LAEs, such as the UV slopes \citep{Santos20} and UV sizes \citep{Malhotra12,Paulino-Afonso18} evolve negligibly over $z=3-6$. This is in line with little evolution in their ionizing properties as the UV slope and UV size are linked to the dust attenuation and SFR surface density \citep[e.g.][]{Reddy21} - properties that likely correlate with \fescs \citep[e.g.][]{Chisholm18}. All this is according to \cite{Harikane2018} finding Ly$\alpha$ escape fractions in LAEs at $z\sim5$ comparable with those of LAEs at $z=2$ using a stacking analysis and H$\alpha$ emission-line measurements inferred from {\it Spitzer}/IRAC colours.

A possible point of confusion is the literature arguing for a strongly evolving \lyas \fescs \citep[e.g.,][]{Hayes2011,Sobral18}. The \lyas \fescs referenced in these works is the ``global" \lyas \fescs averaged over the entire galaxy population. This global \lyas \fescs is computed by dividing the integrated \lyas luminosity density by the integrated UV luminosity density. This quantity indeed grows rapidly with redshift as expected from Fig \ref{fig:intro}. In our framework, the proportion of LAEs among the overall galaxy population increases with redshift, but the average \lyas \fescs of these LAEs is kept fixed to the value measured at $z\approx2$. That is, the rising global \lyas \fescs is a consequence of the increasing LAE fraction.

\section{Results} \label{sec:results}

\begin{figure*}
\centering
\includegraphics[width=0.75\linewidth]{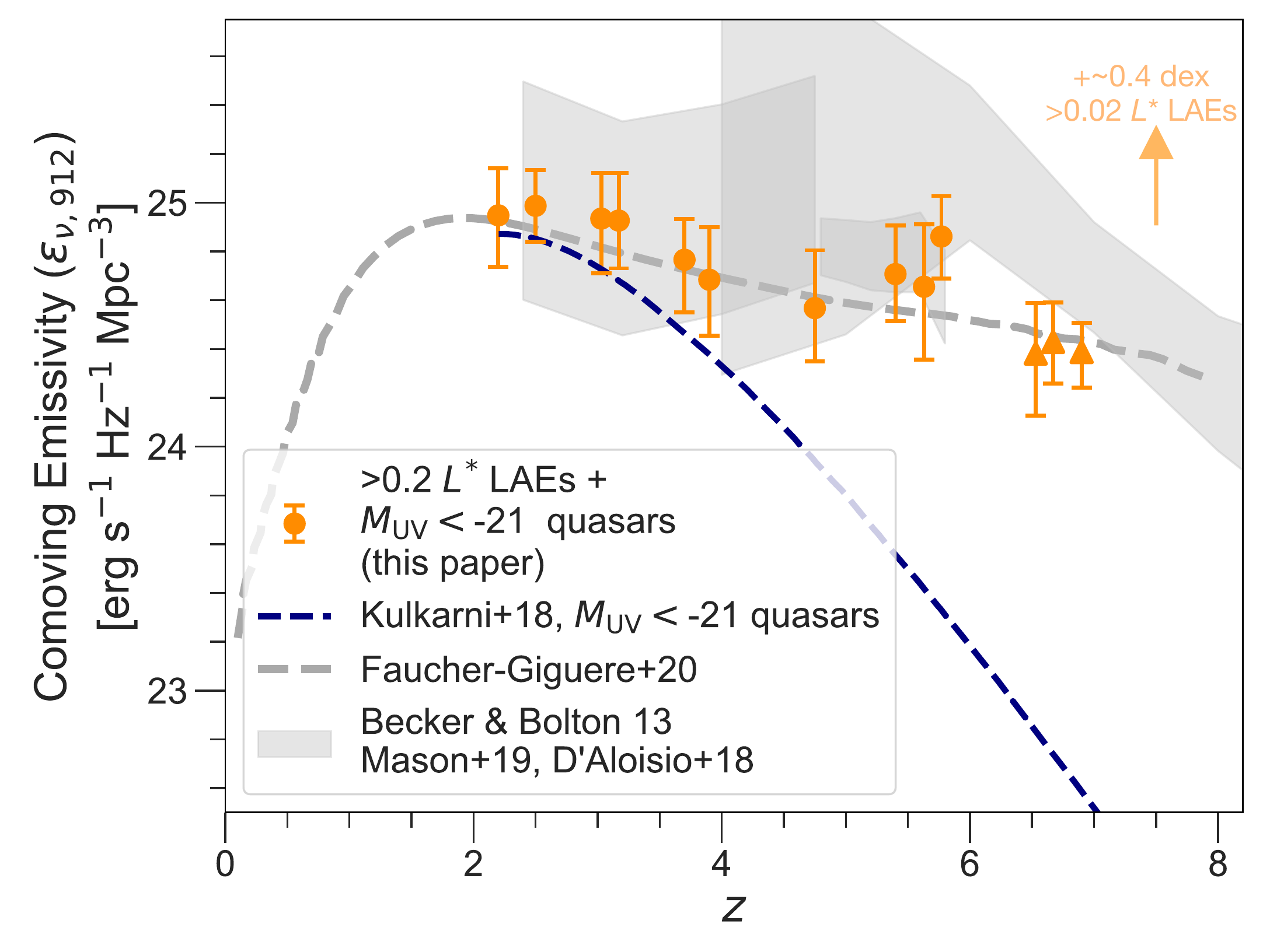}
\caption{Ionizing emissivity as a function of redshift. The orange points show the total emissivity from LAEs (based on Eqn. \ref{eq:nionlyasummary}, Table \ref{tab:knowns}) and quasars for which we assume $f_{\rm{esc}}=1$ (dashed blue line, \citealt{Kulkarni19}). The points at $z=3.1$ and $z=5.7$ have been offset by $\Delta z=0.05$ for clarity. At $z>6$ we use upward pointing triangles to highlight that these measurements are possibly lower limits since the Ly$\alpha$ LFs in the EoR may be attenuated by HI beyond what we assume (Table \ref{tab:knowns}). The orange points are in excellent agreement ($<1\sigma$) with the inferred emissivity at $z>3$ (gray) from the synthesis model of \citealt{FaucherGiguere20} (dashed gray line), the Ly$\alpha$ forest constraints of \citealt{Becker13} ($z\approx2.5-5$) \& \citealt{daloisio18} ($z\approx4.5-6$), and the synthesis model of \citealt{Mason19nonpar} ($z>4$). Bright LAEs dominate the $z>3$ emissivity, and reproduce the gentle evolution of the ionizing background which is at stark odds with the rapidly evolving $\rho_{\rm{SFR}}$ across these redshifts ($\approx$2 dex from $z\approx2-8$, e.g., \citealt{Oesch18}). In the top-right corner we indicate that the contribution from LAEs would move upwards by $\approx0.4$ dex if we assumed the same \lyas \fescs and LyC \fescs hold for $\approx10\times$ fainter LAEs than we have studied. Due to the dominant quasar contribution at $z<3$, this arrow is mostly relevant for $z\approx6$ where LAEs dominate the total emissivity.}
\label{fig:nion}
\end{figure*}

\begin{table}
\centering
\caption{Comoving emissivity  ($\log_{10}{\epsilon_{\rm{\nu,912}}}$) in units of erg s$^{-1}$ Hz$^{-1}$ Mpc$^{-3}$.}
\label{table:emissivity}
\begin{tabular}{lcccc}
\hline
\multicolumn{1}{l}{$z$} &
\multicolumn{1}{c}{Ref.} &
\multicolumn{1}{c}{$>0.2 L^{*}$ LAEs} &
\multicolumn{1}{c}{$>0.2 L^{*}$ LAEs} &
\multicolumn{1}{c}{UVB}\\
\multicolumn{1}{l}{} & \multicolumn{1}{c}{} & \multicolumn{1}{c}{} &
\multicolumn{1}{c}{+ $M_{\rm{UV}}<-21$ AGN} &
\multicolumn{1}{c}{}\\
\hline
\noalign{\smallskip}
2.2 & 1 & $24.15^{+0.19}_{-0.21}$ & $24.95^{+0.19}_{-0.21}$ & 24.92\\
2.5 & 1 & $24.41^{+0.15}_{-0.15}$ & $24.99^{+0.15}_{-0.15}$ & 24.89\\
3.1 & 2 & $24.55^{+0.19}_{-0.22}$ & $24.93^{+0.19}_{-0.22}$ & 24.81\\
3.1 & 1 & $24.53^{+0.19}_{-0.20}$ & $24.93^{+0.19}_{-0.20}$ & 24.81\\
3.7 & 2 & $24.47^{+0.17}_{-0.22}$ & $24.77^{+0.17}_{-0.22}$ & 24.73\\
3.9 & 1 & $24.39^{+0.21}_{-0.23}$ & $24.68^{+0.21}_{-0.23}$ & 24.70\\
4.8 & 1 & $24.45^{+0.24}_{-0.22}$ & $24.57^{+0.24}_{-0.22}$ & 24.61\\
5.4 & 1 & $24.68^{+0.20}_{-0.19}$ & $24.71^{+0.20}_{-0.19}$ & 24.56\\
5.7 & 3 & $24.63^{+0.26}_{-0.30}$ & $24.66^{+0.26}_{-0.30}$ & 24.54\\
5.7 & 4 & $24.85^{+0.17}_{-0.17}$ & $24.86^{+0.17}_{-0.17}$ & 24.54\\
6.6 & 3 & $>24.37^{+0.21}_{-0.25}$ & $>24.38^{+0.21}_{-0.25}$ & 24.46\\
6.6 & 4 & $>24.42^{+0.16}_{-0.17}$ & $>24.43^{+0.16}_{-0.17}$ & 24.46\\
6.9 & 5 & $>24.38^{+0.12}_{-0.14}$ & $>24.39^{+0.12}_{-0.14}$ & 24.44\\
\noalign{\smallskip}
\end{tabular}
\begin{tablenotes}
\item[] References for \lyas luminosity functions: (1) \citet{Sobral18}, (2) \citet{Ouchi2020}, (3) \citet{Konno18}, (4) \citet{Santos16}, (5) \citet{Wold21}. $M_{\rm{UV}}<-21$ AGN LF from \citet{Kulkarni19}. UV background (UVB) values from \citet{FaucherGiguere20}.
\end{tablenotes}
\end{table}

\begin{figure*}
\centering
\includegraphics[width=\linewidth]{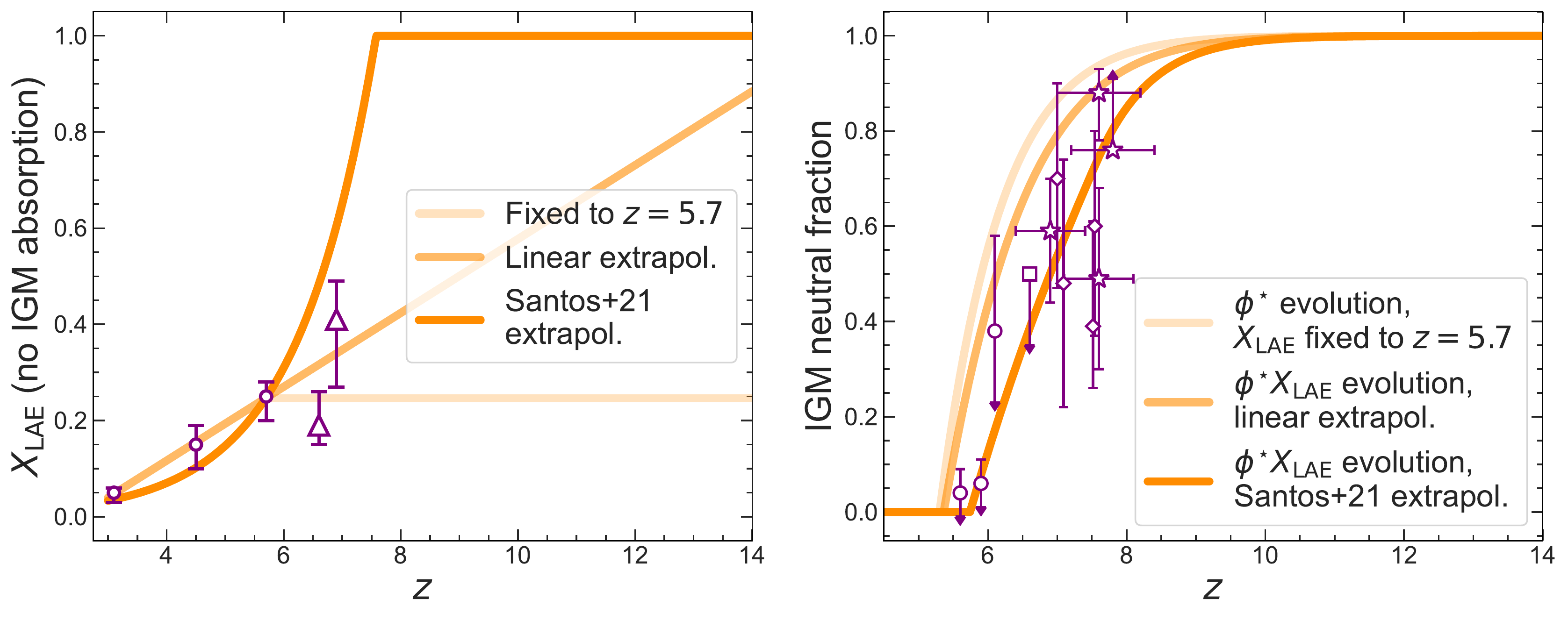}
\caption{\textbf{Left:} Three possibilities for the evolution of the LAE fraction ($X_{\rm{LAE}}$; fraction of $>0.2 L^{*}$ LAEs among $M_{\rm{UV}}<-17$ LBGs) past $z>6$: i) fixed to the value at $z=5.7$ (lightest orange), ii) linearly increasing (orange), iii) non-linearly increasing following \citealt{Santos21} (darkest orange). $X_{\rm{LAE}}$ values inferred from the Ly$\alpha$ LFs listed in Table \ref{tab:knowns} and the framework in \S\ref{sec:LBGframework} are shown as purple points. The $z=6.6$ and $z=6.9$ points based on \citet{Ouchi2020} and \citet{Wold21} are shown with statistical errors, but note that these are likely lower limits due to the expected damping of the Ly$\alpha$ LF during the EoR. \textbf{Right:} Evolution of the IGM neutral fraction ($X_{\rm{HI}}$). The $z>6$ emissivity used to derive $X_{\rm{HI}}$ is obtained by extrapolating the $z=5.7$ emissivity assuming a declining LBG $\phi^{\star}$ \citep{Bouwens21} and an evolving $X_{\rm{LAE}}$ as per the curves in the left panel. The evolution based on \citet{Santos21} produces late and rapid reionization ($90\%$ to $10\%$ between $z\approx8-6$), and is in excellent agreement ($\approx1\sigma$) with literature constraints shown in purple: the dark fraction (circles, \citealt{McGreer15}), LAE clustering (square, \citealt{Sobacchi14}), Ly$\alpha$ damping in LBGs (stars, \citealt{Mason18,Mason19,Hoag19,Jung20}), and quasars (diamonds, \citealt{Davies18a,Yang20,Wang20}).}
\label{fig:xhi}
\end{figure*}

\subsection{LAEs explain the evolution of the ionizing background from $z\approx2-6$} \label{sec:nionexplained}
In Fig. \ref{fig:nion} we present the evolution of the combined quasar and LAE emissivity from our fiducial model at $z\approx2-8$. The quasar emissivity is taken from the fiducial luminosity function of \citet{Kulkarni19} while assuming an \fesc$=1$ and a limiting $M_{\rm UV}<-21$. The quasar and LAE emissivities are comparable at $z\approx2-3$, but at higher redshifts the LAE emissivity dominates the sum (by a factor $\approx100$ at $z\gtrsim6$), see Table $\ref{table:emissivity}$. We note that the estimated LAE emissivities at $z>6$ in Fig. \ref{fig:nion} are depicted as lower limits since the Ly$\alpha$ LFs in the EoR may be significantly attenuated even on the red-side beyond the blue-side corrections we assume (Table \ref{tab:knowns}). We compare this quasar+LAE emissivity to the \textit{total} ionizing emissivity at $z\approx2-8$ from \citet{FaucherGiguere20, Becker13,daloisio18} ($z<6$) and \citet{Mason19nonpar} ($z>6$). The $z\lesssim6$ emissivity is inferred from the opacity of the \lyas forest, while the $z>6$ \citet{Mason19nonpar} emissivity is a non-parametric fit that represents as a summary of recent reionization constraints\footnote{$\dot{n}_{\rm{ion}}(z)$ is allowed to vary by +/- 1 dex in intervals of $\Delta z=1$ from $z=4-14$. The starting point is the \citet{Becker13} emissivity at $z\approx5$ that is incorporated with +/-2 dex errors. The constraints informing the inference include: the \citet{Planck18} Thomson optical depth, the \citet{McGreer15} Ly$\alpha$ forest dark fraction, the clustering of LAEs \citep{Sobacchi14}, the Ly$\alpha$ damping wing of $z>7$ quasars \citep{Davies18a,Greig19} and $z>6$ Ly$\alpha$ EW distributions \citep{Mason18,Mason19,Hoag19}.}. We also show the relative contribution of quasars for comparison.

Combined with the known population of quasars, bright LAEs remarkably match ($<1\sigma$) the normalisation and fairly gentle evolution of $\epsilon_{\nu,912}(z)$ between $z\approx3-6$, without requiring any additional ionizing sources. Matching this evolution is non-trivial. When the emissivity is anchored to $\rho_{\rm{UV}}$, the $\epsilon_{\nu,912}$ increases sharply with cosmic time -- so models that produce reionization by $z=6$ often overshoot the emissivity at $z\approx2-6$ when extrapolated under the same assumptions. A strongly evolving \fescs for unknown or as yet untested reasons is commonly invoked to reproduce the flattening \citep[e.g.,][]{Puchwein19, Finkelstein19,Dayal20,Naidu20, Ocvirk21}. The redshift invariance of LAE LFs over $z=2-6$ thus provides a simple, empirical description of the phenomena plausibly related to the evolution of \fescs. The underlying physics is probably that the correlated set of conditions leading to Ly$\alpha$ (LyC) production and escape become more common with increasing redshift -- i.e., hot stellar populations that blow ionized channels and shine through a dust-free high ionization state ISM (\citetalias{NaiduMatthee22} \citeyear{NaiduMatthee22}) occur in an increasingly large fraction of the galaxy population. We discuss this in more detail in \S $\ref{sec:fescz}$.

In Fig. $\ref{fig:nion}$ we illustrate that the LAE emissivity would increase by $\approx+0.4$ dex if we would assume that the same \lyas \fescs and LyC \fescs are extrapolated to LAEs with luminosities that are 10 times fainter than the LAEs we studied (\citetalias{NaiduMatthee22} \citeyear{NaiduMatthee22}). Due to the large uncertainties in the emissivity constraints, a contribution from such fainter galaxies is therefore not strictly ruled out. We discuss ways to test whether such faint LAEs contribute to the emissivity or not in \S $\ref{sec:roadmap}$. For now, we focus on the results and implications of our fiducial model.
\subsection{LAEs produce late, rapid reionization} \label{sec:reionization}

While the ionizing background at $z<6$ can be directly calculated from Ly$\alpha$ LFs with small IGM corrections, during the EoR these LFs likely only provide lower limits on the emissivity (Fig. \ref{fig:nion}). This is because the neutral IGM damps Ly$\alpha$, extinguishing the line entirely in neutral regions of the Universe \citep[e.g.,][]{Laursen2011,Pentericci14}. The blue-side damping correction we applied at $z<6$ does not sufficiently compensate for this. Further, the LAEs that actually are accounted for in the LFs may be ionizing their own proximity zones (e.g., the $z=6.6$ COLA1 with LyC \fesc$\approx30\%$, \citealt{Matthee18}), so the LyC emitter fraction among the observed $z>6$ LAEs is possibly higher than our adopted $0.5\pm0.1$. Given these effects, we require a model for the evolution of the \textit{intrinsic}, i.e., pre-IGM transmission Ly$\alpha$ LF at these redshifts.

We model the evolution of the Ly$\alpha$ LF into the EoR as an interplay between a sharply falling LBG $\phi^{*}$ and a flat/rising $>0.2 L^{*}$ LAE fraction among these LBGs ($X_{\rm{LAE}}$). This implicitly assumes no evolution in the UV luminosity dependence of the LAE fraction beyond $z>6$. The LBG $\phi^{*}$ is very well constrained over $z=2-10$ \citep{Bouwens21}. The evolution of the intrinsic $X_{\rm{LAE}}$ is however unknown at $z>6$. We illustrate three possible trajectories for $X_{\rm{LAE}}$ in the left panel of Fig. \ref{fig:xhi}: i) fixed to the $z=5.7$ value, ii) a linear extrapolation of the $z\approx2-6$ $X_{\rm{LAE}}$, iii) an exponential extrapolation based on \citet{Santos21} who compared the integrals of the UV and Ly$\alpha$ LFs. Of these, the third trajectory is most favored by the $z=6.6-6.9$ $X_{\rm{LAE}}$ lower limits derived from the \citet{Santos16,Ouchi2020,Wold21} LFs using the simulations described in \S\ref{sec:LBGframework}. This trajectory effectively implies a slight increase in $L^{\star}$ of the intrinsic Ly$\alpha$ LF. The LAE fraction of LAEs is $25^{+4}_{-3}$ \% at $z\approx6$ (consistent with spectroscopic follow-up studies summarised in \citealt{Ouchi2020}), having increased from $\approx5 (15)$\% at $z=3.0 (4.5)$. We cap the LAE fraction to 100\%, which is reached at $z\approx8$.

\begin{table}
\centering
\caption{Comoving emissivity of ionizing photons, $\log(\dot{n}_{\rm{ion}}/\rm{s}^{-1} \rm{Mpc}^{-3})$ arising from $L>0.2 L^{*}$ LAEs under the three assumptions for $X_{\rm{LAE}}$ adopted in \S\ref{sec:reionization} and Figure \ref{fig:xhi}.}
\label{table:nion}
\begin{tabular}{lccc}
\hline
\multicolumn{1}{l}{$z$} &
\multicolumn{1}{c}{Fixed to $z=5.7$} &
\multicolumn{1}{c}{Linear extrapol.} &
\multicolumn{1}{c}{\citet{Santos21} extrapol.}\\
\multicolumn{1}{l}{} & \multicolumn{1}{c}{} & \multicolumn{1}{c}{} &
\multicolumn{1}{c}{}\\
\hline
\noalign{\smallskip}
5.7 & 50.93 & 50.93 & 50.93\\
6.0 & 50.84 & 50.88 & 50.94\\
7.0 & 50.49 & 50.64 & 50.91\\
8.0 & 50.09 & 50.32 & 50.70\\
9.0 & 49.64 & 49.94 & 50.25\\
10.0 & 49.14 & 49.51 & 49.75\\
11.0 & 48.59 & 49.02 & 49.20\\
12.0 & 48.00 & 48.47 & 48.61\\
13.0 & 47.36 & 47.87 & 47.97\\
14.0 & 46.67 & 47.22 & 47.28\\
\noalign{\smallskip}
\end{tabular}
\end{table}

For our three $X_{\rm{LAE}}$ trajectories we present the evolution of the IGM neutral fraction ($x_{\rm{HI}}$) in the right panel of Fig. \ref{fig:xhi}. The emissivity is computed via the framework in \S\ref{sec:formalism}, and the neutral fraction follows from the standard set of reionization equations that balance ionization against recombination \citep[e.g.,][]{Madau99, Robertson13}. For parameters like the clumping factor we assume values adopted in \citet{Naidu20}. The emissivities of the three models are listed in Table $\ref{table:nion}$. The exponentially rising $X_{\rm{LAE}}$ model extrapolated from \citet{Santos21} agrees best ($\lesssim1\sigma$) with the $x_{\rm{HI}}$ literature constraints shown in purple. Note that this was also the only model favored by the $z>6$ $X_{\rm{LAE}}$ lower limits, and hence we designate this our fiducial model. We have also investigated models where $X_{\rm{LAE}}\propto(1+z)^{\alpha}$ with $\alpha=2-7$, and found that these yield similar behaviour as the exponential model, with a preferred $\alpha=5$.

In our fiducial model, the reionization history driven by bright LAEs is late and rapid, with the Universe going from $\approx90\%$ to $10\%$ neutral between $z\approx6-8$, in good agreement with literature constraints that increasingly point to such a rapid timeline (purple points, Fig. \ref{fig:xhi}). As we will show in subsequent sections, the bright ($>0.2 L^{*}$) LAEs occur preferentially in $M_{\rm{UV}}\lesssim-17$ galaxies that are relatively rare and build up sufficient numbers to overpower recombination only at $z\lesssim8$.

\begin{figure*}
\begin{tabular}{cc}
    \includegraphics[width=9cm]{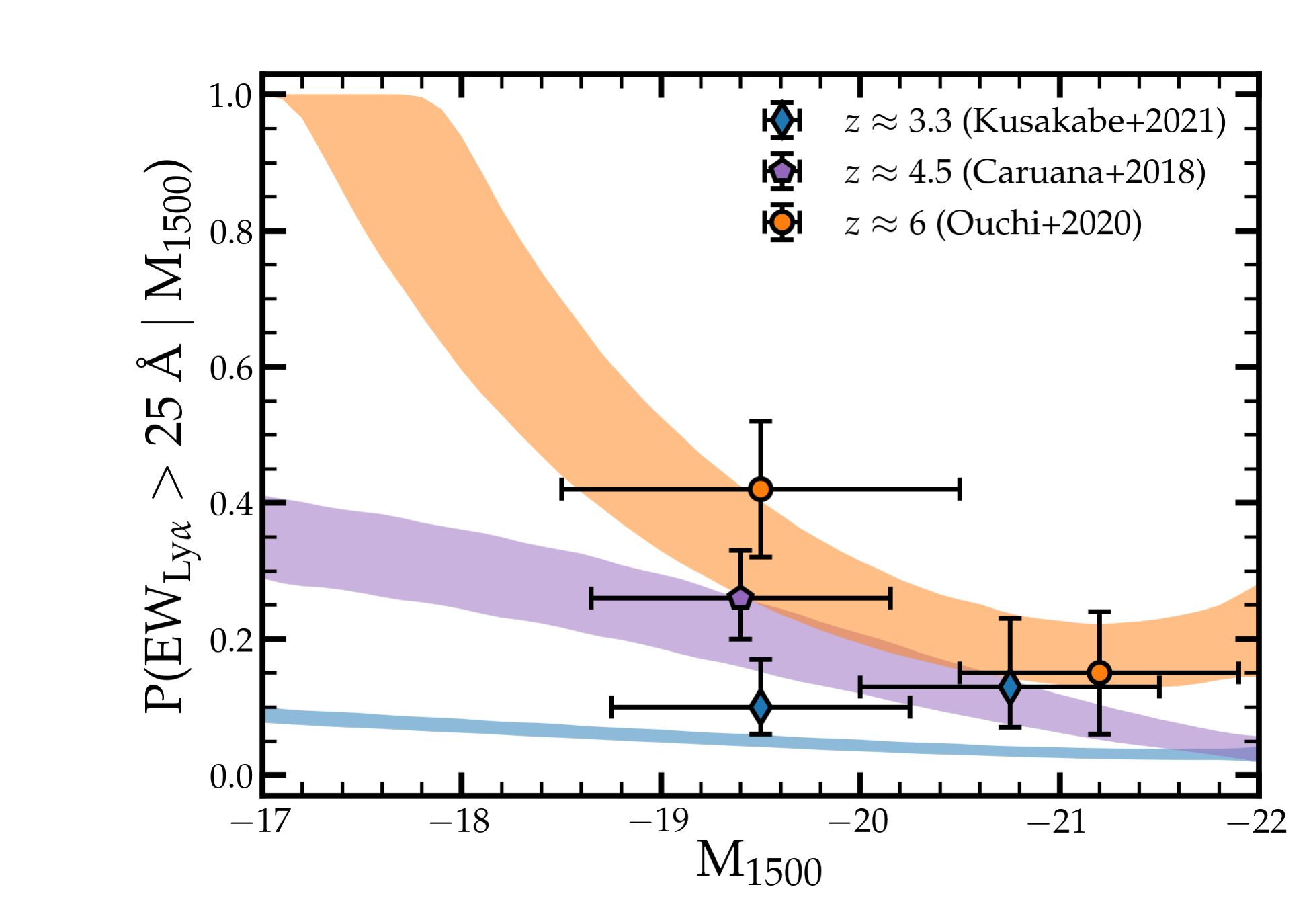} &
\hspace{-0.6cm}    \includegraphics[width=9cm]{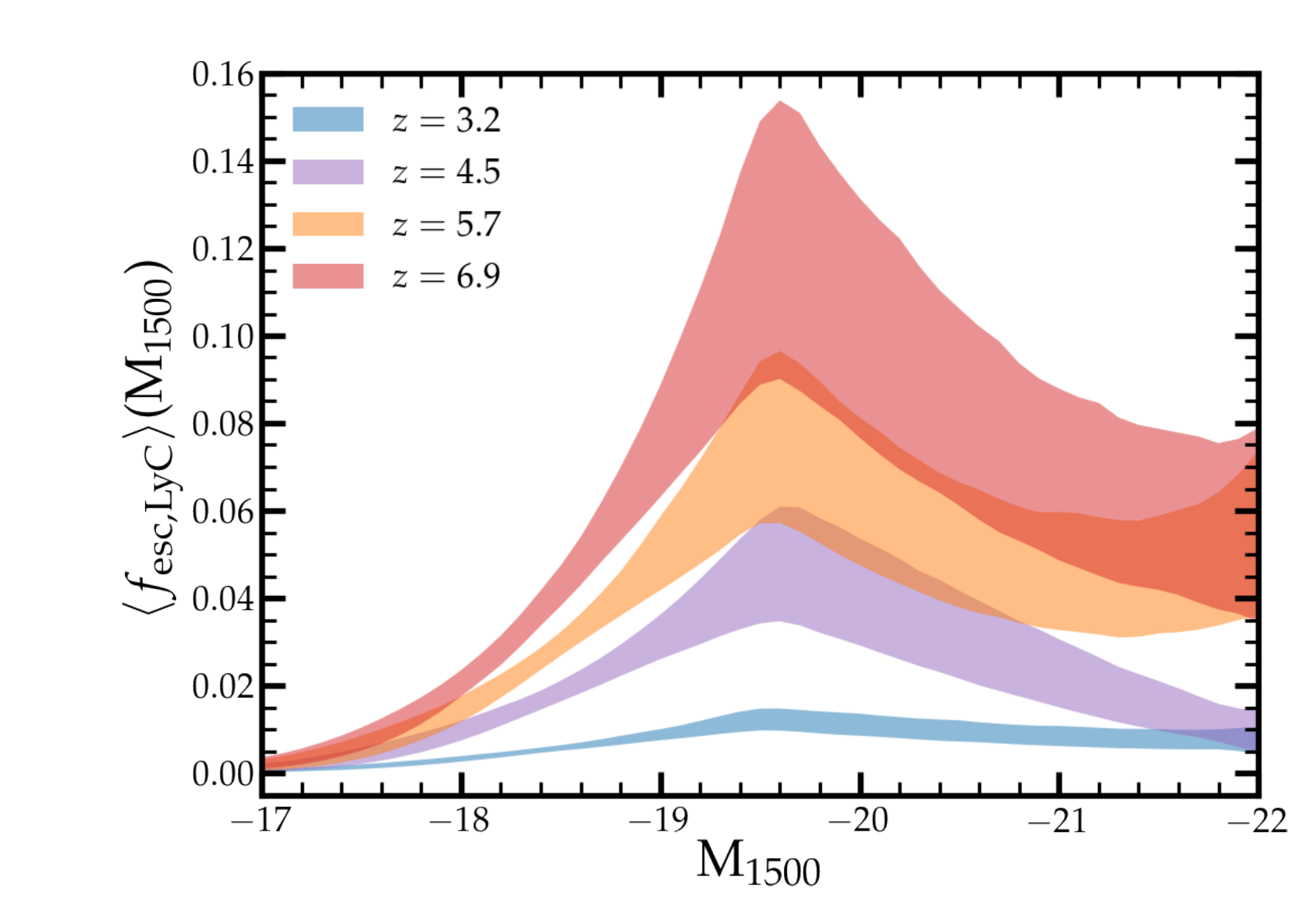} \\
     \includegraphics[width=9cm]{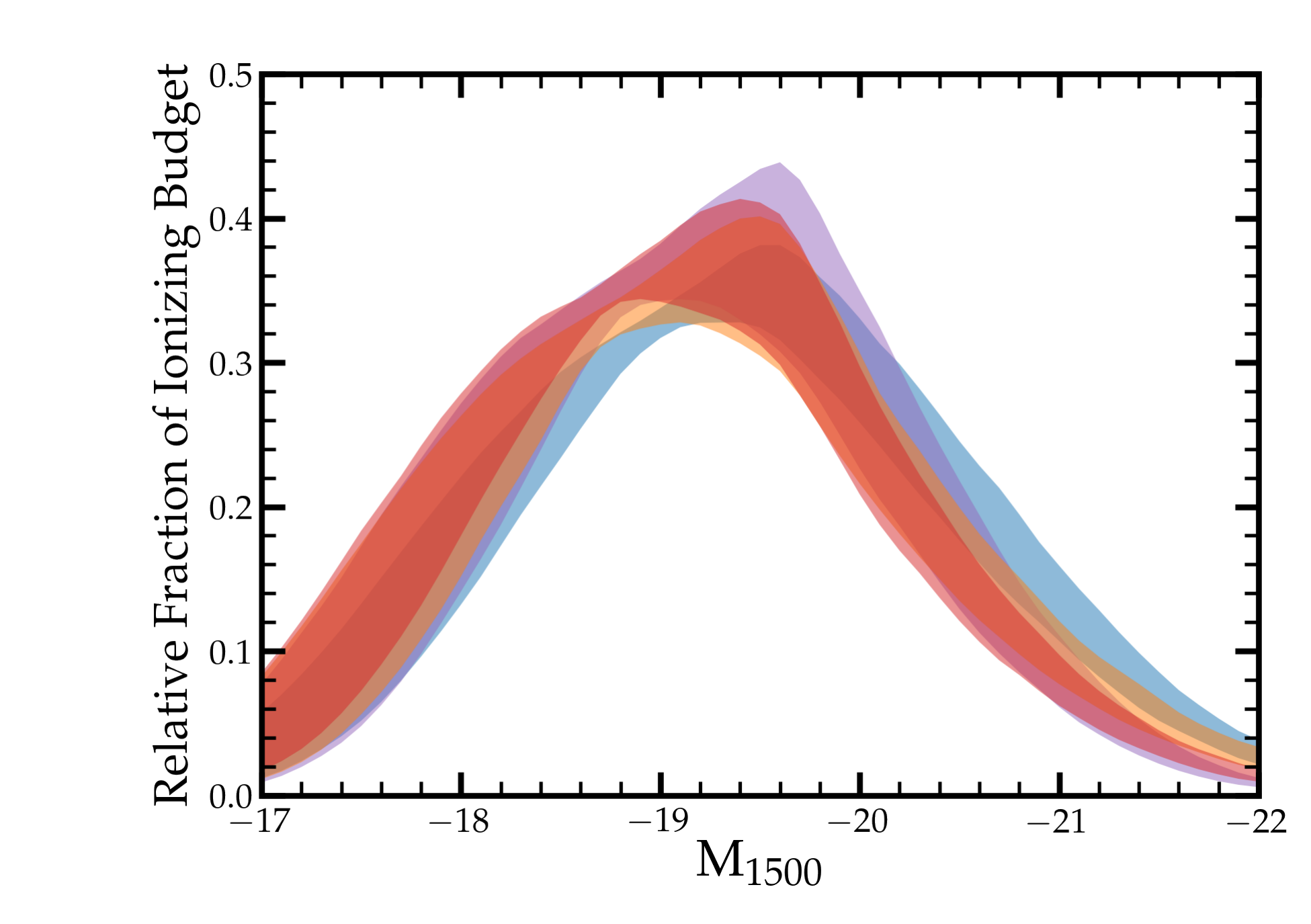} &
\hspace{-0.6cm}    \includegraphics[width=9cm]{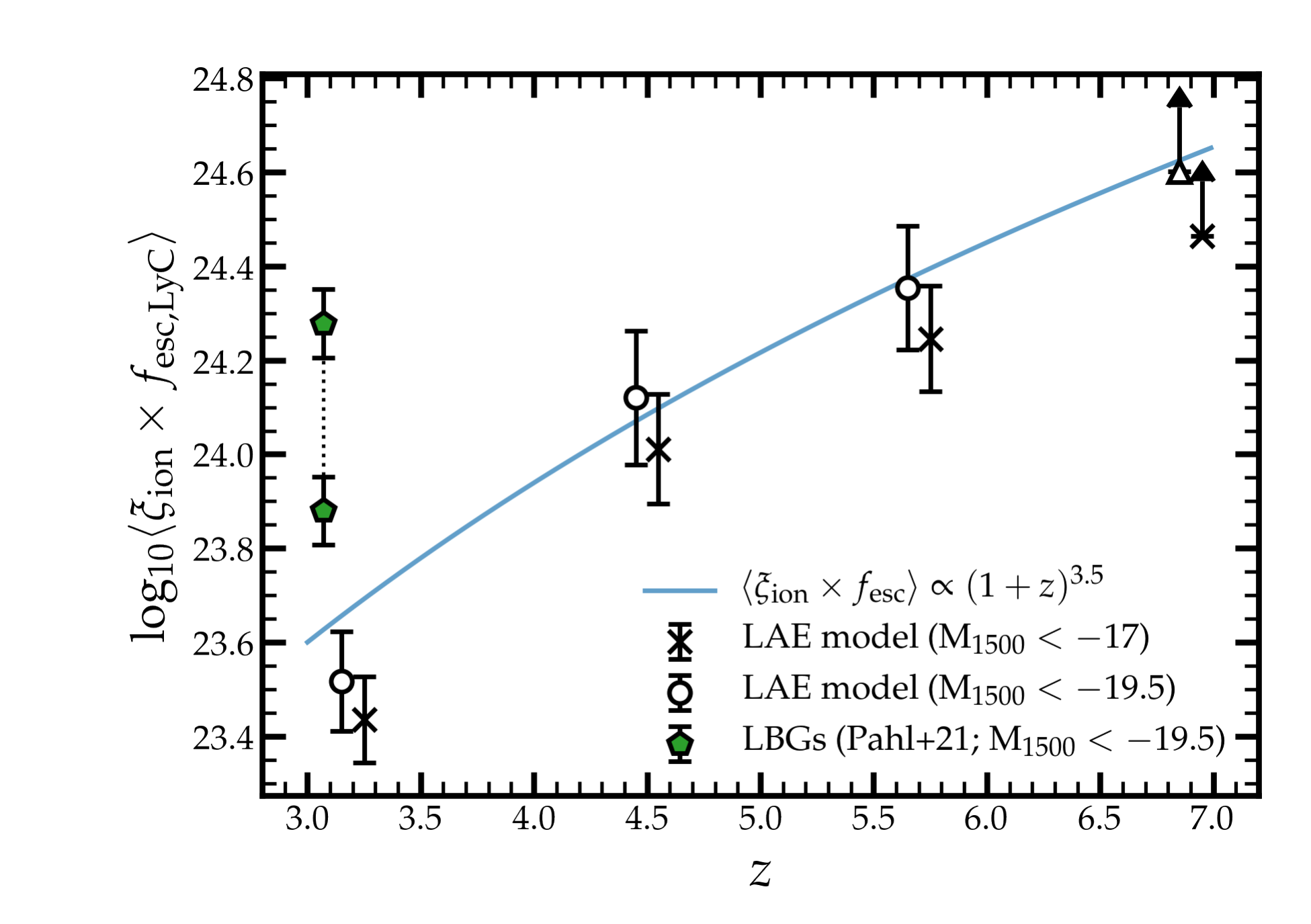} \\
\end{tabular}
\caption{Results from the simulations connecting the LAE-based framework to the LBG-framework (\S $\ref{sec:LBGframework}$). In this simulation, $\langle f_{\rm esc, LAE} \rangle = 25 $ \% for LAEs with luminosity L$_{\rm Ly\alpha}>10^{42.2}$ erg s$^{-1}$ and 0 \% for fainter LAEs. Shaded regions show the 68 \% confidence interval derived from the uncertainties in the parameters used in the simulations. \textbf{Top-left:} the fraction of LAEs (i.e. galaxies with Ly$\alpha$ EW $> 25$ {\AA}) as a function of UV luminosity at redshifts $z=3.2, 4.5, 5.7$ (shaded regions) compared to measurements from \citet{Caruana2018,Kusakabe2020,Ouchi2020}. \textbf{Top-right:} the average escape fraction as a function of UV luminosity from our simulations at redshifts $z=3.2, 4.5, 5.7$ and $z=6.9$. \textbf{Bottom-left:} the relative contribution to the total galaxy emissivity as a function of UV luminosity at the different redshifts. \textbf{Bottom-right:} the average $\langle \xi_{\rm ion} \times f_{\rm esc}\rangle$ for simulated galaxies with M$_{\rm UV}<-19.5$ (open circles) and M$_{\rm UV}<-17$ (black crosses) as a function of redshift. For clarity the points are slightly shifted along the horizontal axis. Data-points at $z>6$ are plotted as lower limits as the Ly$\alpha$ LFs at these redshifts are likely affected by IGM absorption. Our simulations naturally lead to a strong evolution with redshift of $\langle \xi_{\rm ion} \times f_{\rm esc}\rangle \propto (1+z)^{3.5}$. For comparison we show the measurement from \citet{Pahl21} based on a stack of LBGs with M$_{\rm UV}<-19.5$ at $z\approx3$ (green pentagon). The lowered green pentagon shows the results of correcting the average $f_{\rm esc}$ by a factor three due to a bias in galaxies with high Ly$\alpha$ EWs in their parent sample (see Appendix \ref{sec:Pahl}).}
    \label{fig:LAE_model}
\end{figure*}

\section{Connection to the full galaxy population} \label{sec:LBGframework}
It is of interest to directly compare the results from the LAE emissitivity to the UV luminosity formalism. For example, in the LBG-framework, the relevant \fescs is $\langle f_{\rm esc, LBG} \rangle$, i.e. \fescs averaged over the population of UV-selected galaxies. In \S\ref{sec:results}, we showed that we can match the emissivity evolution using LAEs when $\langle f_{\rm esc, LAE} \rangle = 25 $ \% for LAEs with luminosity L$_{\rm Ly\alpha}>10^{42.2}$ erg s$^{-1}$ and 0 \% for fainter LAEs, independent of redshift. However, in order to relate this averaged \fescs to an average for the full UV-selected population we need to take the evolving LAE fraction (and the UV luminosity distribution of LAEs) into account. 

\subsection{Statistically connecting LAEs \& LBGs using simulations}
We obtain the connection of our LAE-based emissivity model to a LBG-based formalism using a simple set of simulations. We generate separate populations of LAEs and LBGs in an average region of the universe with a volume of $5\times10^8$ cMpc$^3$ at the redshifts $z=3.2, 4.5, 5.7, 6.9$ following the LAE luminosity functions by \cite{Konno18,Sobral18,Herenz19,Wold21} and UV luminosity functions by \cite{Bouwens21} -- both parametrised with Schechter functions -- down to a limiting Ly$\alpha$ luminosity of $10^{41}$ erg s$^{-1}$ and a UV luminosity of $-15$, respectively. The Ly$\alpha$ luminosities are increased by factors 1, 1.07, 1.17 and 1.17 at these redshifts which correct for the average impact of the IGM on the blue side of the Ly$\alpha$ line of LAEs \citep{Hayes21}, ignoring any potential additional impact from reionisation at $z=6.9$ \citep[e.g.][]{Wold21}. We obtain the distribution of UV luminosities for the simulated LAEs based on a Ly$\alpha$ EW distribution and a UV slope. The EW distribution of LAEs is assumed to be invariant of redshift or luminosity and follows $n({\rm EW})\propto \exp{(-{\rm EW/EW_0})}$, where ${\rm EW}_0=90\pm30$ {\AA} \citep[e.g.][]{Hashimoto2017,Santos20,Kerutt2022}. By selection, the minimum possible EW for each LAEs is 25 {\AA}. The UV slope is fixed to $\beta=-2.1\pm0.2$. This step allows us to statistically derive the LAE fraction for LBGs as a function of UV luminosity by simply counting the number of LAEs of a given UV luminosity in the simulated universe and comparing this to the number of LBGs with this luminosity. The LAE fractions derived here are internally consistent with the population-averaged LAE fraction ($X_{\rm LAE}$) discussed in \S $\ref{sec:reionization}$. Note that our simple model is agnostic of the Ly$\alpha$ EW distribution {\it of LBGs} for Ly$\alpha$ lines with EW $<25$ {\AA}: whether these have Ly$\alpha$ in absorption or how the lower EWs are distributed simply does not impact the model.

We then obtain the average \fescs in each UV luminosity bin by counting the fraction of LAEs with a luminosity above L$_{\rm Ly\alpha}>10^{42.2}$ erg s$^{-1}$ (the minimum {\it output} Ly$\alpha$ luminosity of leaking galaxies within our fiducial model), and multiplying this fraction by the average escape fraction of LAEs with this luminosity, i.e. $\langle f_{\rm esc}\rangle=25\pm5$ \% in the fiducial model. We also calculate the average ionizing photon production efficiency, $\xi_{\rm{ion}}$, in a similar way, where we assign $\xi_{\rm{ion}}=10^{25.74}$ Hz erg$^{-1}$ to the fraction of LBGs that are LAEs (based on H$\alpha$ and UV luminosity measurements; \citetalias{NaiduMatthee22} \citeyear{NaiduMatthee22}) and $\xi_{\rm ion}=10^{25.5}$ Hz erg$^{-1}$ for the remaining fraction of LBGs based on detailed fits to the rest-frame UV stellar spectra \citep{Steidel18}. Similarly, we also calculate the emissivity-weighted $\langle f_{\rm esc} \, \xi_{\rm ion} \rangle$ for all LBGs with $>0.2 L^{\star}$ luminosity and the relative contribution to the emissivity from LAEs with various UV luminosities. At each redshift, we perform 10,000 simulations while perturbing the input parameters with their uncertainties. The results from these simulations are shown in Fig. $\ref{fig:LAE_model}$. In Appendix $\ref{sec:modelvar}$ we show and discuss the results from various model variations.

\subsection{Results from simulations}
The top-left panel of Fig. $\ref{fig:LAE_model}$ shows the LAE fractions among LBGs in our simulations compared to results from spectroscopic follow-up studies of LBGs. The modeled LAE fractions are only sensitive to the EW distribution and the relative shapes of the UV and LAE LFs. While our model is very simple (i.e. it does not assume variations in EWs or UV slopes for LAEs with luminosity or redshift), it quantitatively reproduces the observed redshift evolution of the LAE fraction and the UV-luminosity dependence of the LAE fraction at $z\approx6$, which provides a useful model-validation. 

Our model suggests that, on average, $f_{\rm esc}$ is highest at a UV luminosity of M$_{\rm UV}\approx-19.5$, or $\approx0.25 L^{\star}$ (Fig. $\ref{fig:LAE_model}$, top-right). The increase from M$_{\rm UV}=-17$ to $M_{\rm UV}=-19.5$ arises due to the minimum Ly$\alpha$ luminosity floor of the galaxies that have $f_{\rm esc}>0$. Galaxies with M$_{\rm UV}\approx-17$ require extremely rare Ly$\alpha$ EW in order to reach the Ly$\alpha$ luminosity floor, while more common EWs are sufficient for brighter galaxies. The decrease at higher luminosities is a consequence of the decreasing fraction of LAEs (and therefore LyC leakers) with increasing UV luminosity. The low escape fractions at $z\approx3$ are in good agreement with upper limits obtained in e.g. \cite{Grazian17}. The ionising output from galaxies is proportional to the luminosity function, galaxies' line-luminosity and their escape fraction. In the bottom-left panel of Fig. $\ref{fig:LAE_model}$ we show that this results in a roughly log-normal distribution of the luminosity dependence of the relative contributions to the total emissivity with a clear peak around M$_{\rm UV}\approx-19.5$. Compared to the UV luminosity dependence of the escape fraction (top-right panel of Fig. $\ref{fig:LAE_model}$), the higher importance of fainter galaxies in the distribution of the budget is due to their higher number densities. 
The emissivity distribution depends mostly on the luminosity floor and is thus almost redshift-invariant in our model. The slight shift of the distribution towards a relatively higher contribution from faint galaxies at higher redshifts is primarily due to the steepening of the faint-end slope of the UV luminosity function with redshift \citep[e.g.][]{Bouwens21}.

Finally, in the bottom-right panel of Fig. $\ref{fig:LAE_model}$ we show the average $\langle \xi_{\rm ion} \times f_{\rm esc}\rangle$ for LBGs with UV luminosities brighter than M$_{\rm UV} = -19.5$ and $-17$ as a function of redshift. These averages are weighted by the UV luminosity dependent contribution to the emissivity (i.e. the bottom-left panel). The average for $M_{\rm UV}<-19.5$ allows a direct comparison to the results from \cite{Pahl21} (see also \citealt{Steidel18}) that we discuss in Appendix $\ref{sec:Pahl}$. We find a clear increase of the produced ionising photons that escape with redshift, roughly following $\langle \xi_{\rm ion} \times f_{\rm esc}\rangle \propto (1+z)^{3.5}$. This is very similar to the required evolution in the escape fraction in order to reconcile the UV background with the cosmic star formation rate density \citep[e.g.][]{HaardtMadau2012}. The majority of this evolution can be attributed to evolution in the average escape fraction (i.e. $\propto(1+z)^{\alpha}$ where $\alpha\approx3$) with only a modest increase in the average $\xi_{\rm ion}$ over $z=3-7$. This is similar to what is seen for the {\it LBG population-averaged} Ly$\alpha$ escape fraction \citep{Hayes2011,Sobral18}. The average LyC escape fraction $\langle f_{\rm esc} \rangle$ among $M_{\rm UV}<-17$ galaxies increases from $1.1\pm0.3$ \% at $z=3.2$ to $7.1\pm1.8$ \% at $z\approx7$. This population-averaged escape fraction is likely underestimated at $z>6$ due to the increased impact of the IGM on the Ly$\alpha$ observability. In case the $z\approx7$ Ly$\alpha$ LF is underestimated due to incomplete reionization, we find that $\langle f_{\rm esc} \rangle$ could increase to $13.8\pm3.0$ \%, when we would use the fiducial extrapolation of the $X_{\rm LAE}$ discussed in \S $\ref{sec:reionization}$. An $\langle f_{\rm esc} \rangle$ of $13.8\pm3.0$ \% may seem low compared to the literature consensus of $\sim20$ \% for M$_{\rm UV}<-13$ galaxies to complete reionization \citep[e.g.][]{Robertson15}, and particularly also because our fiducial model considers relatively brighter galaxies. However, note that the population averaged $\xi_{\rm ion}$ in our fiducial model at $z\approx7$ is $10^{25.75}$ Hz erg$^{-1}$, which is significantly higher than canonical assumptions of $10^{25.2}$ Hz erg$^{-1}$, mitigating a lower escape fraction.

\begin{figure*}
\centering
\includegraphics[width=\linewidth]{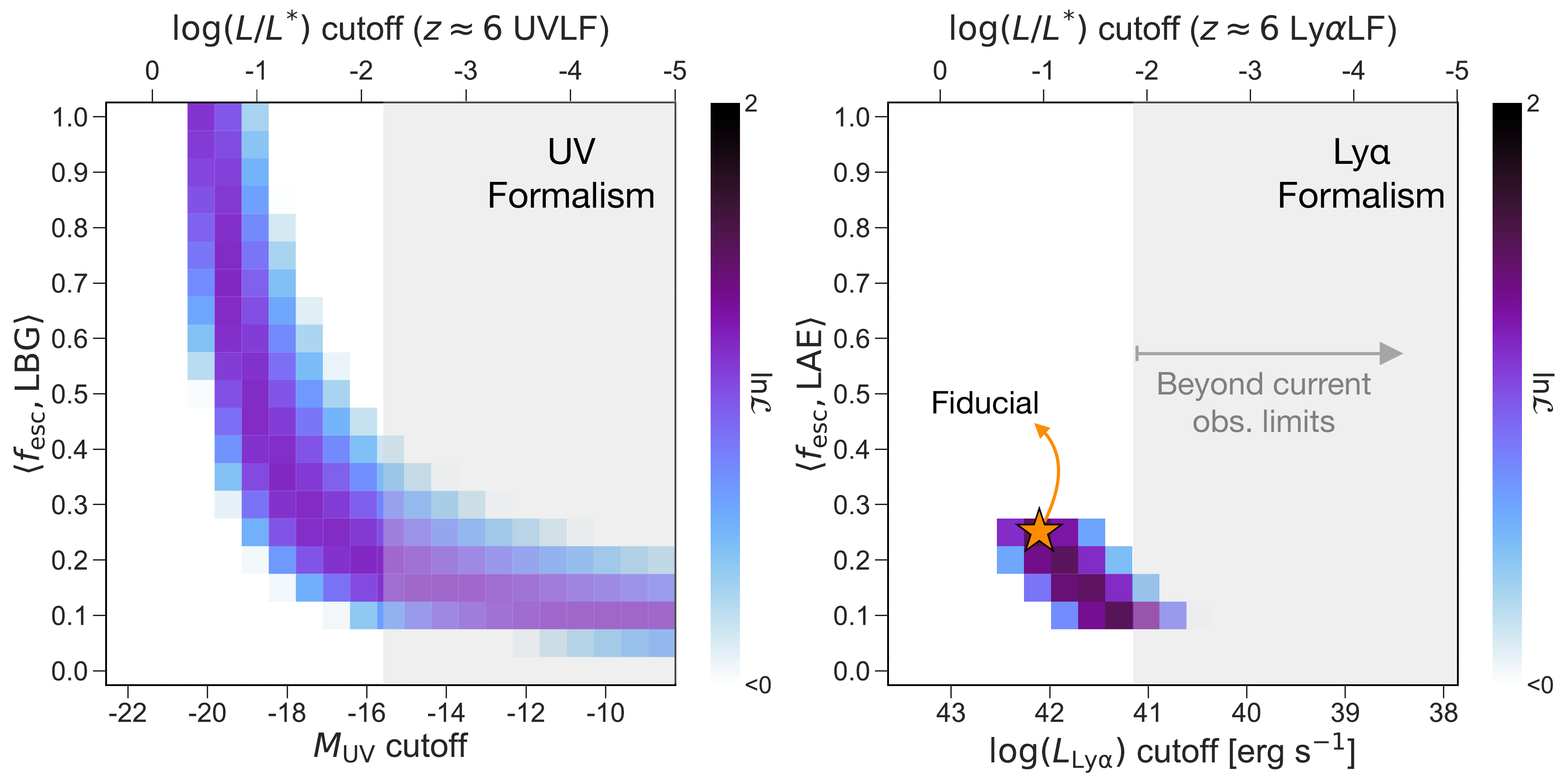}
\caption{Comparison of the degeneracy between the LyC \fescs and integration depth in the UV formalism (\textbf{left}) and Ly$\alpha$ formalism (\textbf{right}). As an illustrative example we evaluate the log likelihood, ln$(\mathcal{L})$, of various $\langle$\fesc$\rangle$ and integration depth combinations against the $z\approx6$ $\epsilon_{\rm{912}}$ from \citet{daloisio18}. The colors in the plot correspond to the likelihood of combinations, with darker colors representing more likely combinations. We assume $\log_{10}(\xi_{\rm{ion}}$/Hz erg$^{-1})=25.2$ for the UV formalism \citep[e.g.,][]{Robertson15}, and the values in Table \ref{tab:knowns} for the Ly$\alpha$ formalism. In the UV formalism one may need to integrate the luminosity function and assume ionizing properties for galaxies much fainter than those that are currently observed, while this is not the case for the Ly$\alpha$ formalism (see e.g. the sensitivities from \citealt{Bacon21} and \citealt{delavieuville19}).}
\label{fig:formalisms}
\end{figure*}

\section{Discussion} \label{sec:discussion}

\subsection{The LAE-emissivity model and \fesc(z) of LBGs} \label{sec:fescz}

The need for an evolving galaxy population-averaged LyC \fescs among LBGs has long been realized due to the low average \fesc$\approx0-5\%$ inferred at $z\approx1-3$ \citep[e.g.][]{Siana10,Grazian16,Grazian17,Rutkowski16,Steidel18} and the requirement for a population-averaged \fesc$\approx20\%$ at $z>6$ under the canonical ionizing photon production efficiency $\log_{10}(\xi_{\rm{ion}}$/Hz erg$^{-1})=25.2$ \citep[e.g.,][]{HaardtMadau2012,Robertson15, Bouwens15c,Naidu20,FaucherGiguere20}. Qualitatively, strong redshift evolution of $\langle f_{\rm esc} \rangle$ over this redshift interval may not be unexpected due to significant changes in the dust content \citep{Bouwens16}, galaxy sizes (and star formation rate surface density; e.g. \citealt{Shibuya15,Naidu20}), the burstiness of star formation \citep{Faisst16b,Faisst2019, Tacchella20} and ionisation parameter \citep{Sanders2020}. All these properties, to some extent, may correlate with \fescs \citep[e.g.][]{Heckman11,Faisst16,Vasei16,Izotov18b,Izotov21,Chisholm18,Chisholm20}, but so far it has been challenging to identify clear relations between such properties and \fesc.
Strong evolution in $\langle f_{\rm esc} \rangle$ is also expected in case \fescs would be significantly higher in very low mass galaxies \citep[e.g.][]{Katz18,Hutter21,Ocvirk21} as their relative share of the integrated UV luminosity increases towards higher redshifts due to the steepening of the faint-end slope of the UV LF. However, in order to reconcile the high escape fractions required to reionise the Universe with the evolution of the UVB after reionisation, one requires reionisation to be self-inhibiting, i.e. the main sources of reionisation to shut down directly due to the completion of reionization itself \citep[e.g.][]{Finkelstein19,Ocvirk21} or indirectly due to a side-effect such as dust production.

With our LAE-based framework we argue that the relative evolution of the Ly$\alpha$ luminosity density and the UV luminosity density (e.g. Fig. $\ref{fig:intro}$) is an outcome of the same underlying phenomena that regulate the evolution of the population averaged \fesc. This follows from the intimate connection between escaping Ly$\alpha$ and LyC luminosities. As there are indications that the population averaged \fescs for LAEs does not evolve (as indicated by the redshift-invariance of average Ly$\alpha$ line-profiles; \citealt{Hayes21}), the strong observed increase in the LAE fraction \citep[e.g.][]{Stark11,Kusakabe2020} with redshift therefore directly implies a strong evolution in the average \fescs of the full LBG population. 

The fact that the LAE emissivity model matches the required $\langle f_{\rm esc}\rangle$ evolution does not mean it explains why this happens. Rather, it reduces the problem to one that is more tractable: explaining the evolution of the LAE fraction among galaxies. The Ly$\alpha$ output from galaxies may vary due to differences in Ly$\alpha$ production and escape \citep[e.g.][]{Sobral18,Trainor19,Runnholm20}. It is plausible that both vary strongly with cosmic time: the build-up of dust may increases the covering factor of dense and obscured sight-lines that prohibit the likelihood of observing a galaxy as a LAE \citep[e.g.][]{Atek2008,Henry15,RiveraThorsen15,Matthee16,Yang17,Trainor19} and young metal-poor stellar populations with high ionising photon production efficiency become less common at lower redshift. In fact, in our companion paper (\citetalias{NaiduMatthee22} \citeyear{NaiduMatthee22}) we showed that the various conditions promoting a high Ly$\alpha$ (and LyC) output are correlated. The galaxies with high inferred \fescs show highly ionizing stellar populations capable of powering nebular \CIV\ and \HeII\ emission that shine through a dust-free, porous ISM. LAEs with a low inferred \fescs on the other hand have somewhat older stars and a more dusty, less ionised and covered ISM. If the mechanisms controlling the production and escape of Ly$\alpha$ and LyC photons occur in concordance it is not unlikely that the evolution is as strong as $\propto(1+z)^3$.

A more quantitative assessment of whether we can {\it explain} the evolution of the LAE fraction (and thus $\langle f_{\rm esc}(z)\rangle$) requires us to predict the Ly$\alpha$ luminosity output based on a suite of multi-variate correlations with properties such as absorption line strengths, optical emission-line ratios, dust attenuation and SFR surface density \citep[e.g.][]{Trainor19,Runnholm20}. It would be appealing to use such multi-variate correlations to match the Ly$\alpha$ LF and EW distribution based on the UV luminosity function and knowledge of the distributions of the relevant properties among the galaxy populations at $z>3$. This is currently unfeasible given the poorly constrained properties of the ISM in high-redshift galaxy samples, but should be attempted in the future. Once one is able to match the evolution of the Ly$\alpha$ LF, our framework suggests that such models also match the evolution of the globally-averaged \fescs in galaxies.

A final caveat needs to be stated, which is that it is possible that the ionizing properties of the population of LAEs themselves may evolve. As we discussed, the invariance of average Ly$\alpha$ profiles over $z=2-6$ \citep{Hayes21,Matthee21} suggests that the HI column density does not vary significantly, such that the average \fescs in LAEs likely evolves little. Further, the inferred Ly$\alpha$ escape fraction in LAEs at $z\approx5$ is comparable to the average at $z=2$ \citep{Harikane2018}. However, it is possible that the Ly$\alpha$ escape fraction varies (at fixed \fesc) in case there are significant differences in the dust-to-gas ratio, possibly due to differences in gas-phase metallicity. It is also possible that the relative distribution of leaking and non-leaking LAEs changes. Additionally, it is possible that the intrinsic Ly$\alpha$ EW evolves with redshift, for example due to a varying stellar metallicity. If such changes are present, it is expected that lower metallicities at higher redshifts yield higher Ly$\alpha$ escape fractions and EWs. As shown in \S $\ref{sec:formalism}$, a change in Ly$\alpha$ \fescs could lead to a lower emissivity than estimated in our fiducial model. Higher EWs do not impact the LAE-based emissivity, but they do impact the translation into the UV luminosities associated to the LAEs (\S $\ref{sec:LBGframework}$ and Appendix $\ref{sec:modelvar}$). Therefore, if intrinsic EWs are higher at high-redshift, the UV luminosity of the optimal ionizers may decrease slightly. 

Unlike the LyC \fesc, the Ly$\alpha$ escape fraction can be directly measured at $z=3-6$ in the near future with plausibly only a modest impact of the IGM (as discussed in \S $\ref{sec:knowns}$). Million galaxy \lyas surveys in this redshift range are already underway (e.g., HETDEX, \citealt{Gebhardt21}). Rest-frame optical observations of faint LAEs that combine {\it JWST} and VLT/MUSE spectroscopy could directly extend the measurements of the (variation among) Ly$\alpha$ line-profiles (with known systemic redshifts) and Ly$\alpha$ escape fraction into these uncharted regimes and test these model assumptions.

\subsection{Equivalence with canonical reionization budget} \label{sec:equivalence}
Here we compare the total ionizing budget in the EoR derived using the Ly$\alpha$-based framework to canonical UV-anchored calculations \citep[e.g.,][]{Robertson15, duncan&conselice15, Ishigaki18}. The key difference is how this budget is distributed -- while the canonical approach assumes a fixed \fescs and a $\xi_{\rm{ion}}$ across all galaxies that contribute in proportion to their UV luminosity, the fiducial implementation of our bright LAE framework isolates a small subset of highly efficient ionizers.

Consider the emissivity at $z\approx7$, a redshift at which most models and the \citet{Planck18} $\tau$ suggest reionization is underway. Integrating the \citet{Bouwens21} UV LF down to $M_{\rm{UV}}=-13.5$ as is typically done under standard assumptions (\fesc$=0.2$, $\log({\xi_{\rm{ion}}/\rm{Hz\ erg^{-1}}})\approx25.2$, e.g.,  \citealt[][]{Robertson13,Robertson15,Mason15}) produces an ionizing budget approximately equally split between $M_{\rm{UV}}<-17$ and $M_{\rm{UV}}>-17$ galaxies.

We are interested in $M_{\rm{UV}}=-17$ since in our fiducial calculation there is almost no contribution from galaxies fainter than this limit (bottom-left, Fig. \ref{fig:LAE_model}) -- the faint galaxies' integrated UV luminosity comprising $\approx50\%$ of the total is ``unproductive'' and immaterial to the ionizing emissivity. Further, only a fraction ($25$ to $50\%$) of the brighter $M_{\rm{UV}}<-17$ galaxies are LAEs (Figs. \ref{fig:xhi} and $\ref{fig:LAE_model}$), and of these only $\approx50\%$ are LyC leakers. That is, galaxies that are contributing only $\approx10\%$ of the integrated UV luminosity ($50\% \times 25\ \rm{to}\ 50\% \times 50\%$) likely produce the entire ionizing budget in the fiducial model (resembling the `oligarchs' in \citealt{Naidu20}). However, these galaxies have a high production and escape of ionizing photons ($\log({\xi_{\rm{ion}}/\rm{Hz\ erg^{-1}}})\approx25.9$, \fesc$\approx50\%$, \citetalias{NaiduMatthee22} \citeyear{NaiduMatthee22}), and are thus able to produce the required emissivity with a combined \fesc$\times\xi_{\rm{ion}}$ that is $>10\times$ higher than canonical models. We note that despite this being a back-of-the-envelope calculation, the presented numbers agree well with the detailed simulations in \S\ref{sec:LBGframework} that account for subtleties such as e.g., the detailed luminosity-dependence of $X_{\rm{LAE}}$. 

\begin{figure}
\centering
\hspace{-1cm}    \includegraphics[width=9.2cm]{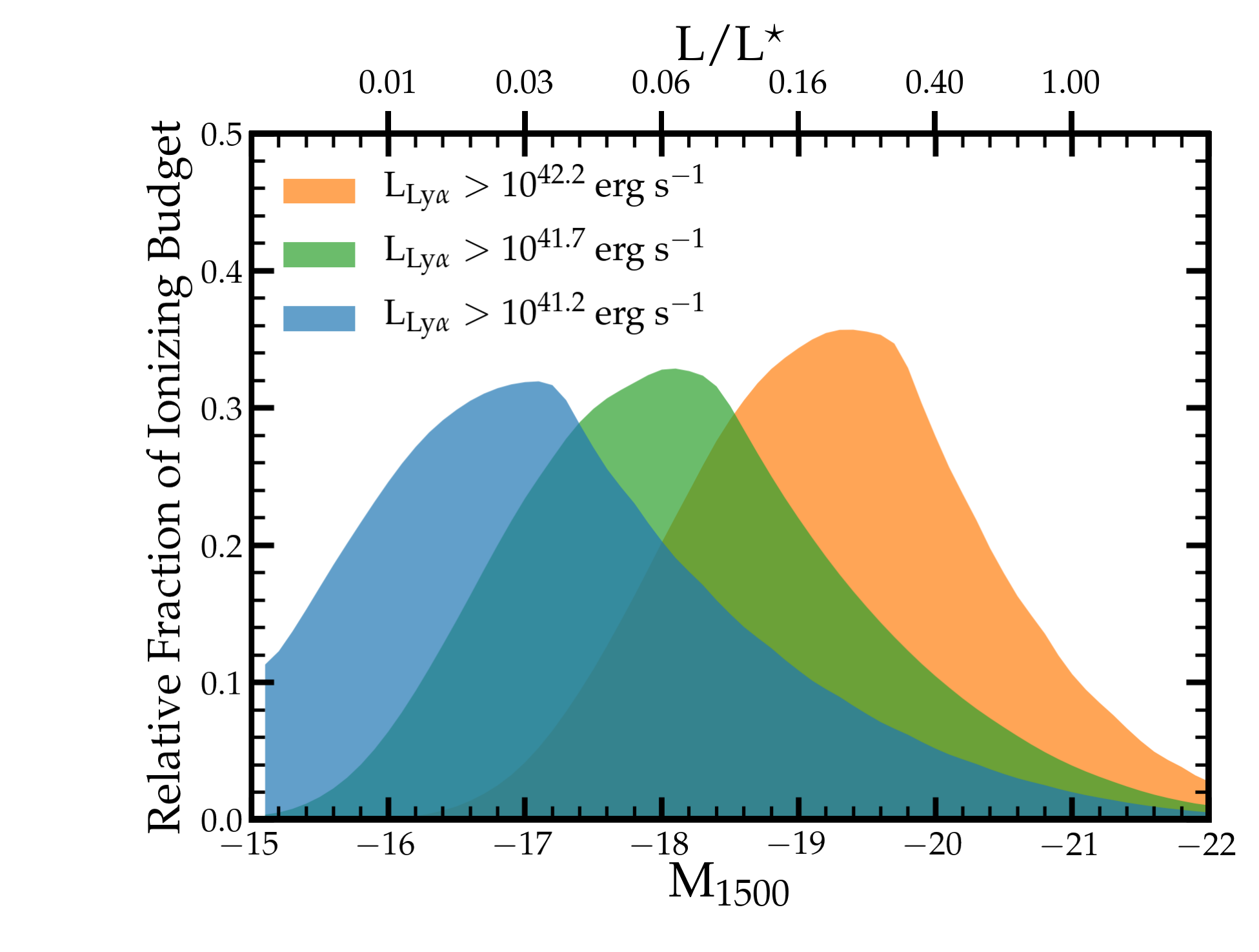}\\
\caption{Distribution of the ionising budget as a function of the UV luminosity in the LAE-emissivity model with a fixed \fescs for LAEs at $z=5.7$ (as described in \S $\ref{sec:LBGframework}$). The orange curve shows the fiducial model where the limiting Ly$\alpha$ luminosity of the faintest ioniser is $10^{42.2}$ erg s$^{-1}$. Green and blue curves show when fainter LAEs contribute to the emissivity. Except, for the cut-off luminosity, the Ly$\alpha$ and LyC \fescs of LAEs are independent of luminosity in these models. }
\label{fig:budget}
\end{figure}

To generalise the results of the LAE-emissivity model, we explore the consequences of the degeneracy between the luminosity limit and the average escape fraction in order to achieve a similar total ionizing emissivity.\footnote{We note that, as shown in Eq. $\ref{eq:nionlyasummary}$, changes in the Ly$\alpha$ escape fraction could also impact the LAE-emissivity. However, the Ly$\alpha$ escape fraction has been more directly measured for the relevant galaxies as presented in \S $\ref{sec:knowns}$.} In Fig. $\ref{fig:formalisms}$, we show this degeneracy using both the UV and LAE formalisms at $z\approx6$. The key difference between the allowed parameter range in the formalisms is that we can cap the range of allowed escape fractions for LAEs based on our results in \citetalias{NaiduMatthee22} (\citeyear{NaiduMatthee22}), and that the escape fraction of LAEs is higher than that of LBGs such that the regime with the strongest degeneracy is ruled out. Furthermore, the ionizing photon production efficiencies of LAEs are higher than the canonically assumed value for LBGs \citep[e.g.][]{Sobral2017}. Because $\dot{n}_{\mathrm{ion, LAE}}(z)$ has a stronger dependency on the escape fraction than $\dot{n}_{\mathrm{ion, LBG}}(z)$, the degeneracy between the escape fraction and the limiting luminosity is less important than it is in the case for the UV formalism. When ten times fainter LAEs are allowed to contribute to the emissivity, the LAE emissivity model yields quite a different distribution of the ionizing budget (see Fig. $\ref{fig:budget}$ and Appendix $\ref{sec:modelvar}$). In such a scenario, $\approx80$ \% of the galaxies with $M_{\rm{UV}}>-17$ are active ionisers and the faintest sources that contribute to the ionising budget have $M_{\rm{UV}}\approx-14$, which is more comparable to `democratic' reionization \citep[e.g.][]{Finkelstein19}. We discuss how to possibly distinguish these scenarios in the LAE-framework next.

\subsection{Which galaxies dominate the emissivity and how to test the key parameters?} \label{sec:lowerlimit} \label{sec:roadmap}
In the fiducial model, the galaxies that contribute to the emissivity have UV luminosities $\approx0.025-2.5\times$ L$^{\star}$ and the emissivity is dominated by relatively typical ($\approx0.25$ L$^{\star}$) galaxies (see \S $\ref{sec:LBGframework}$). These UV luminosities correspond to stellar masses of $\approx10^7-2\times10^9$ M$_{\odot}$ \citep[e.g.][]{Song16}, with the highest escape fraction found for galaxies with a mass $\approx2\times10^8$ M$_{\odot}$. Interestingly, the approximate mass for which we find that \fescs peaks in the fiducial model is very similar to the mass with maximum \fescs in the recent simulations from \cite{Ma20}. These authors show that at lower masses, inefficient star formation and feedback are unable to clear low-column density channels through which ionising photons can escape, while dust attenuation decreases \fescs at higher masses \citep[see also][]{Cen20}.

As we show in Fig. $\ref{fig:budget}$ (discussed further in Appendix $\ref{sec:modelvar}$), the limiting Ly$\alpha$ luminosity of the faintest ioniser has significant impact on the UV luminosity distribution of the ionising budget. For example, in case the limiting Ly$\alpha$ luminosity is ten times fainter, very faint $\approx0.03$ L$^{\star}$ galaxies would dominate the emissivity. In that case, the typical stellar mass of the dominant ionisers would decrease to $\approx10^7$ M$_{\odot}$. As discussed above and shown in Fig. $\ref{fig:formalisms}$, the escape fraction of LAEs is strongly related to the luminosity of the faintest source that contributes to the emissivity. For this reason, it is crucial to further test whether the indirectly inferred population-averaged escape fraction of our representative sample of LAEs is indeed $\langle f_{\rm esc}\rangle = 25$ \% (\citetalias{NaiduMatthee22} \citeyear{NaiduMatthee22}) with direct (stacking) of LyC measurements \citep[e.g.][]{Pahl21}. This would require deep UV imaging data from {\it HST}/WFC3 at $z\approx2$ \citep[e.g.][]{Oesch18b}. Unfortunately, this is in practice very challenging for our sample because the source-density of these LAEs is relatively low and their UV continuum relatively faint, requiring very deep UV observations over several {\it HST} pointings.

The emissivity constraints have significant uncertainties \citep[e.g.][]{Becker13,Gallego21} and the exact value of the \fescs required to reionise the Universe is uncertain. For example, the mean free path of ionizing photons could be shorter than previously thought, which would mean the required ionizing emissivity should be higher \citep[e.g.][]{Davies21,Cain21}. Therefore, perhaps a more stressing question to address is whether the LyC escape fraction of faint galaxies differs from that of brighter galaxies. It is feasible to indirectly infer the luminosity dependence of \fescs using Ly$\alpha$ line-profile statistics as a function of Ly$\alpha$ luminosity. This requires observations of gravitationally lensed, intrinsically faint LAEs that extend to ten times fainter luminosities than the XLS-$z2$ sample. In our fiducial model, we expect the Ly$\alpha$ line-profiles of fainter LAEs to have larger peak separations and little to no Ly$\alpha$ photons emerging around the systemic redshift compared to brighter LAEs. In addition to a potentially different \fesc, it is also possible that fainter LAEs have a different Ly$\alpha$ escape fraction. This can also be directly tested with joint measurements of the Ly$\alpha$, H$\beta$ and H$\alpha$ lines \citep[e.g.][]{Matthee21}. By establishing whether the Ly$\alpha$ and (inferred) LyC escape fractions vary between LAEs with different emerging Ly$\alpha$ luminosity, we will be able to verify the fiducial model presented in this paper, or warrant the inclusion of fainter LAEs to the ionizing budget.

\subsection{A `Disco' UV background?} \label{sec:disco}
A key feature of our fiducial model is that a minority of galaxies produce the emissivity (half the bright LAEs with $M_{\rm{UV}}<-17$). Further, \citetalias{NaiduMatthee22} (\citeyear{NaiduMatthee22}) suggest that the timescale when a galaxy leaks LyC is on the order of $10$ Myrs, matched to the lifetimes of the most ionising stellar populations. This picture is supported by latest hydrodynamical simulations that emphasize the stochastic, bursty nature of \fescs  \citep[e.g.,][]{Trebitsch17,Rosdahl18,Kimm19,Ma20,Barrow20}. Furthermore, the escape of ionising (and Ly$\alpha$) photons is likely strongly directionally dependent \citep[e.g.][]{Behrens2014,Zheng2014,Fletcher19,Smith2019}. The cocktail of spatial rarity, temporal stochasticity, and directional bias results in an ionizing background comprised of rapid flashes arising from more or less random locations in more or less random directions.

At $z\gtrsim5$, when LAEs dominate the emissivity (Fig. \ref{fig:nion}), one might expect the ionizing background to thus be ``disco-like". We speculate that this could contribute to the Ly$\alpha$ opacity fluctuations observed in high-redshift quasar spectra \citep[e.g.][]{Becker15,Bosman18,Bosman21,Eilers18,Kulkarni19b,Keating20,Bosman21}, and possibly enhancements in the ionising background in the vicinity of metal absorbers \citep[e.g.][]{Finlator16}. Dedicated simulations that incorporate such a stochastic background will help test this hypothesis. It would be interesting to explore local enhancements in the emissivity around LAEs using Ly$\alpha$ transmission statistics \citep[e.g.][]{Kakiichi18,Meyer20} separating samples by their Ly$\alpha$ line-profile. Such a test should be performed in the post-reionization era in order to mitigate the impact of a possibly reduced Ly$\alpha$ damping wing transmission.

The stochasticity may also prove a discriminator for whether bright or faint galaxies dominate the ionizing background. While in the fiducial model, $\approx25\%$ of the $M_{\rm{UV}}<-17$ galaxies dominate the emissivity (e.g. \S $\ref{sec:equivalence}$), in the faint model galaxies extending to far fainter $M_{\rm{UV}}<-15$ contribute (see bottom panel of Fig. \ref{fig:modelvar}). When multitudes of faint galaxies dominate the emissivity, the ionizing background is likely steady because the temporal and spatial stochasticity is smoothed out over an enormous number of sources, i.e., a ``floodlight" ionizing background where numerous tiny bulbs unite to shine in a stable manner, as opposed to a UV background dominated by rare bright sources that can yield large-scale opacity fluctuations \citep[e.g.][]{Chardin17}.

\subsection{Testing the IGM-unaffected Ly$\alpha$ Fraction ($X_{\rm{LAE}}$) at $z>6$}
A key uncertainty in our framework is the integrated Ly$\alpha$ luminosity density at $z\gtrsim6$ (Fig. \ref{fig:xhi}) due to the possibly large IGM corrections. This uncertainty impacts the timeline of reionization in the LAE-framework. It is possible to directly measure the fraction of bright LAEs among $z>6$ galaxies during the EoR through Ly$\alpha$ spectroscopy or narrowband imaging of the largest ionized bubbles as more of them come into view (e.g., the one around EGSz8p7 at $z\approx8.7$, \citealt{Leonova21}, \citealt{Naidu21JW}, \citealt{Tilvi20}). The Ly$\alpha$ photons escaping galaxies within large-scale ionized bubbles will be able to redshift out of resonance before encountering an IGM damping wing, thus making the measurement of the IGM-unaffected Ly$\alpha$ fraction feasible \citep[e.g.,][]{Mason20,Qin21}. Large proximity zones around high-$z$ quasars will allow for a similar measurement -- a prototype for this kind of observation is the $z\approx5.7$ quasar and three LAEs around it presented in \cite{Bosman20}. The bright LAE `Aerith B' in this study shows a Ly$\alpha$ line profile suggestive of a very low \fesc, which suggests that there is significant scatter in \fescs among LAEs and calls for larger statistical samples.

\section{Summary}
In this paper we present a model to account for the relatively mild evolution of the cosmic ionizing emissivity from star-forming galaxies over $z=2-8$ based on Ly$\alpha$ emitters (LAEs, \S $\ref{sec:formalism}$). We argue that LAEs are the natural subset of the galaxy population that is responsible for the ionizing background in addition to quasars. This follows from the intuitive and empirically proven connection between Ly$\alpha$ \fescs and LyC \fesc, and the strong ionising properties of LAE stellar populations. 

We base our {\it fiducial} calculation of the maximal LAE-emissivity on the \fescs and $\xi_{\rm{ion}}$ of bright LAEs ($L_{\rm{Ly\alpha}}>0.2 L^{*}$) at $z\approx2$ obtained in the companion paper (\citetalias{NaiduMatthee22} \citeyear{NaiduMatthee22}), where we argued half these LAEs have \fesc$\approx50\%$, $\log(\xi_{\rm{ion}}$/Hz erg$^{-1})\approx25.9$, while the other half do not significantly contribute to the emissivity. We assume fainter LAEs and non-LAE LBGs make no contributions either, but discuss the implications if fainter galaxies would contribute. We argue for redshift-invariance of the ionizing properties of such bright LAEs based on the observed invariance of several key properties (e.g., $\Sigma_{\rm{SFR}}$, sizes, $\beta_{\rm{UV}}$), and particularly their Ly$\alpha$ profiles. These properties are plausibly linked to the ionizing output of a galaxy. As indicated from the strong differences in the evolution of the Ly$\alpha$ and UV luminosity densities over $z=2-6$, the LAE-ionizing population is a rare fraction of the total galaxy population at $z\approx2-3$, but forms a significant subset of galaxies at $z>6$. [\S $\ref{sec:formalism}$-\S $\ref{sec:knowns}$, Fig. $\ref{fig:intro}$ and Table $\ref{tab:knowns}$].

We show that: 
\begin{itemize}
    \item The combined emissivity of bright LAEs and quasars can reproduce the relatively flat observed emissivity at $z\approx2-6$. This result is non trivial given that both the star-formation density and quasar number density increase from $z\approx6$ to $z\approx2$ and because we do not fine-tune the integration limit down to which luminosity galaxies contribute. [\S $\ref{sec:nionexplained}$, Fig. $\ref{fig:nion}$]
    
    \item Also, the LAEs in the fiducial model produce late and rapid reionization between $z\approx6-9$ with negligible contribution from quasars and a plausible extrapolation of the relative evolution of the Ly$\alpha$ and UV luminosity density into the EoR. [\S $\ref{sec:reionization}$, Fig. $\ref{fig:xhi}$]
    
    \item Connecting the LAEs to the general galaxy population, we show that the LAE emissivity model model naturally yields a rise in (population-averaged) LyC \fescs from $\approx1\%$ at $z\approx3$ to $\gtrsim10\%$ at $z>7$ and this is explained entirely by the growing Ly$\alpha$ emitter fraction with redshift that the model reproduces. The averaged ionising efficiency increases in concert. [\S $\ref{sec:LBGframework}$, Fig. $\ref{fig:LAE_model}$]
    
    \item The LAE emissivity model naturally produces a peak in the relation between UV luminosity and \fescs at $\approx0.15$ L$^{\star}$ ($M_{\rm{UV}}\approx-19$), suggesting that the majority of ionising photons originate from galaxies with M$_{\star}\approx2\times10^8$ M$_{\odot}$. This is in agreement with recent simulations that predict an optimum mass for \fescs as massive galaxies are dusty, while less massive galaxies have poor star-formation efficiencies. [\S $\ref{sec:LBGframework}$, Fig. $\ref{fig:LAE_model}$]
    
    \item In the fiducial model, a highly ionizing minority of $M_{\rm{UV}
    }<-17$ star-forming galaxies -- contributing only $\approx10$ \% of the integrated UV luminosity at $M_{\rm{UV}}<-13.5$ -- is responsible for the entirety of the reionization budget, with a combined \fesc$\times\xi_{\rm{ion}}$ that is $\approx10\times$ higher than canonical models. [\S $\ref{sec:equivalence}$]

    \item  The fiducial model yields a UV background that is dominated by rare galaxies with relatively bright stochastic flashes of LyC that produce a ``disco" ionizing background with significant temporal and spatial fluctuations, in particular at $z\approx5-6$ where galaxies dominate significantly over quasars. We speculate that this may contribute to the large observed opacity variations in high-redshift quasar spectra. [\S $\ref{sec:disco}$]
    
\end{itemize}

Our simple fiducial model - only half the LAEs with a luminosity $>0.2 L^{\star}$ contribute to the emissivity and other star-forming galaxies do not - works remarkably well in producing the right emissivity and, non-trivially, its evolution, an optimum in the \fescs - M$_{\rm UV}$ relation, and the redshift evolution of the population averaged \fesc. The Ly$\alpha$ luminosity range of the ionising sources in our model has been fully probed by current surveys, therefore the galaxies are known to exist, their properties have been constrained, and their number densities are measured. 

We discuss various caveats and the impact of assumptions in \S $\ref{sec:discussion}$. The most significant of these are the role of fainter LAEs and whether the ionizing properties of LAEs vary with luminosity or redshift. Any Ly$\alpha$-luminosity limited emissivity model will yield a similar redshift evolution of the ionizing emissivity and thus account for strong redshift evolution in the global \fescs of the galaxy population over $z\approx2-8$. The LAE emissivity model generally yields a natural peak in the relation between the escape fraction and UV lumionosity, and thus a roughly log-normal distribution of the ionizing budget with luminosity and plausibly mass. The position of the peak depends mostly on the luminosity limit of the faintest ionizer. As we show in Appendix $\ref{sec:modelvar}$, in case LAEs with ten times fainter luminosities would contribute, the population averaged \fescs should be a factor two lower in order to match the $z\approx6$ emissivity. In such a model, the emissivity would be dominated by relatively common faint galaxies with $M_{1500} \approx -17$ and the UV background will be more uniform. We propose that resolved Ly$\alpha$ line-profile observations extending the XLS-$z2$ survey \citep{Matthee21} to fainter luminosities can distinguish these models. In the fiducial model, the Ly$\alpha$ profiles of faint LAEs should be broader with larger peak separations and less flux at the systemic velocity, compared to brighter LAEs, while the Ly$\alpha$ profiles of faint LAEs and bright LAEs should be comparable if faint galaxies had a dominant role. Redshift evolution of the various model parameters can likewise be tested when {\it JWST} enables the joint study of Ly$\alpha$ and rest-frame optical lines at $z>3$. The LAE framework therefore opens up the prospects of empirically testing whether the cosmic emissivity has been dominated by numerous low-mass galaxies or rarer, more massive galaxies.

\section*{Acknowledgements}
We thank an anonymous referee for an encouraging and constructive report that helped improving the quality of this work. We acknowledge illuminating conversations with Xiaohan Wu, Chris Cain, Anna-Christina Eilers, Simon Lilly and Ruari Mackenzie.
RPN gratefully acknowledges an Ashford Fellowship granted by Harvard University. MG was supported by NASA through the NASA Hubble Fellowship grant HST-HF2-51409. PO acknowledges support from the Swiss National Science Foundation through the SNSF Professorship grant 190079.
GP acknowledges support from the Netherlands Research School for Astronomy (NOVA). MH is fellow of the Knut and Alice Wallenberg Foundation. DE is supported by the US National Science Foundation (NSF) through Astronomy \& Astrophysics grant AST-1909198.
The Cosmic Dawn Center (DAWN) is funded by the Danish National Research Foundation under grant No.\ 140. RA acknowledges support from Fondecyt Regular Grant 1202007. ST is supported by the 2021 Research Fund 1.210134.01 of UNIST (Ulsan National Institute of Science \& Technology).
MLl acknowledges support from the ANID/Scholarship Program/Doctorado Nacional/2019-21191036.
JC acknowledges support from the Spanish Ministry of Science and Innovation, project PID2019-107408GB-C43 (ESTALLIDOS) and from Gobierno de Canarias through EU FEDER funding, project PID2020010050.

\section*{Data availability}
The data underlying this article were accessed from the ESO archive. The raw ESO data can be accessed through http://archive.eso.org/cms.html. The derived data generated in this research will be shared on reasonable request to the corresponding authors.

\bibliographystyle{mnras}
\bibliography{MasterBiblio.bib}

\appendix

\section{Impact of variations on model parameters} \label{sec:modelvar}
In the main text, we mainly focused on the results from our fiducial LAE-emissivity model in which only the contributions from LAEs were included that are brighter than $>0.2$ L$^{\star}$. This is because the Ly$\alpha$ and LyC escape fractions have been probed directly for these LAEs (\citetalias{NaiduMatthee22} \citeyear{NaiduMatthee22}). The contribution from bright LAEs alone suffices in matching the emissivity level and its evolution over $z=2-6$ (in particular the emissivity level recently published by \citealt{FaucherGiguere20}, see Fig. $\ref{fig:nion}$). However, as discussed in \S $\ref{sec:nionexplained}$, the inclusion of ten times fainter LAEs (with fixed escape fractions) is still (marginally) consistent with the uncertainty on various emissivity constraints at $z=2-8$ (Fig. $\ref{fig:nion}$). Moreover, LAEs can match the rapidness of the reionization of the Universe with a milder extrapolation of their relative abundances compared to LBGs at $z>6$ (\S $\ref{sec:reionization}$).

\begin{figure}
\centering
\hspace{-1cm}    \includegraphics[width=9.2cm]{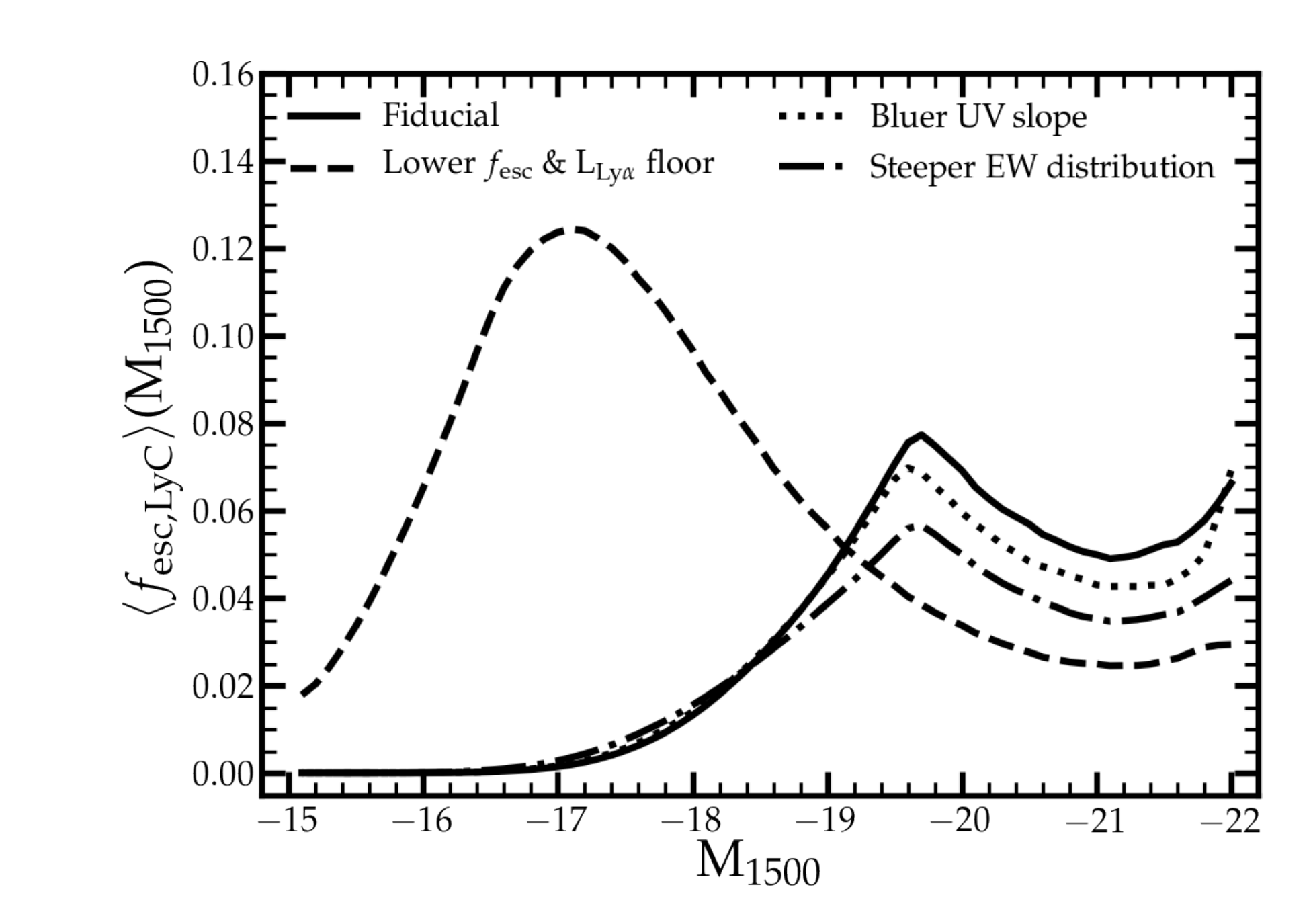}\\
\hspace{-1cm}    \includegraphics[width=9.2cm]{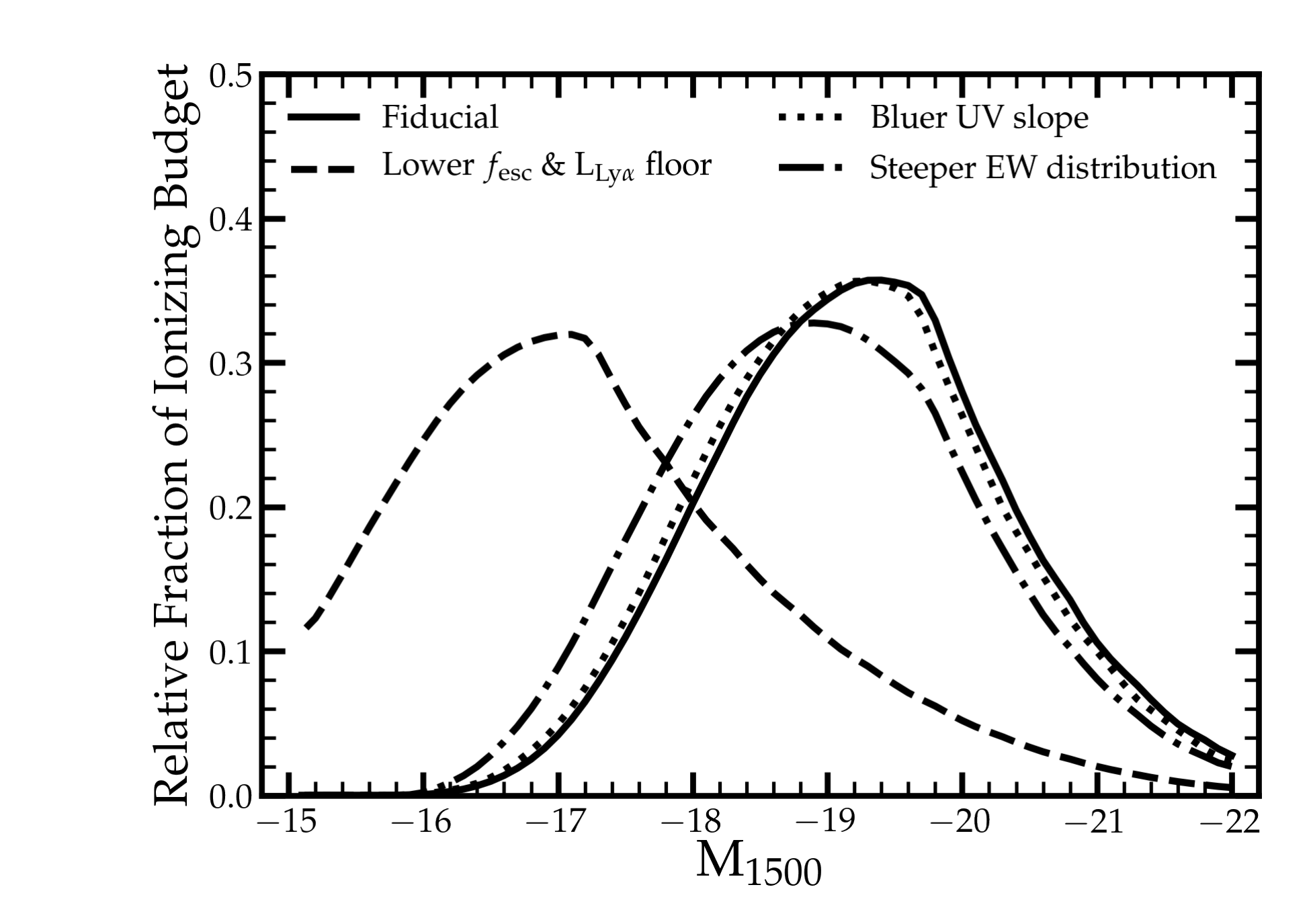}\\
\caption{The effect of model variations on the average \fesc (top) and the fractional contribution to the ionising budget (bottom) as a function of UV luminosity. We show results at $z=5.7$ but note that the relative differences to the fiducial model are not strongly sensitive to the specific redshift. The solid line shows our fiducial model ($\langle f_{\rm esc} \rangle$ = 25 \% for L$_{\rm Ly\alpha}>10^{42.2}$ erg s$^{-1}$, EW$_0$=90 {\AA} and $\beta=-2.1$), while the dotted lines shows the results for $\beta=-2.3$, the dot-dashed line for EW$_0$=135 {\AA} and the dashed line the results for a model where $\langle f_{\rm esc} \rangle$ = 12.5 \% for L$_{\rm Ly\alpha}>10^{41.2}$ erg s$^{-1}$.  }
\label{fig:modelvar}
\end{figure}

In Fig. $\ref{fig:modelvar}$, we illustrate how various choices in the LAE-emissivity model and its connection to the LBG population impact the UV luminosity dependence of the escape fraction (top panel) and how the ionizing budget is distributed (bottom panel). We focus on the redshift $z=5.7$, but note that the relative changes between model-choices are not strongly dependent on the specific redshift. The following variations are investigated: 1) changing the UV slope of LAEs ($\beta=-2.3$ instead of $-2.1$), 2) changing the slope of the exponential EW distribution of LAEs from 90 to 135 {\AA} (this implies a similar increase in the typical EW of LAEs and is within the range of published scale lengths; see e.g. \citealt{Hashimoto2017}) and 3) including the ionizing contribution of up to 10 times fainter LAEs (integrating LFs down to $10^{41.2}$ erg s$^{-1}$ with a population averaged $\langle f_{\rm esc}\rangle=12.5$ \% in order to match the total emissivity).

Fig. $\ref{fig:modelvar}$ shows that the translation between the LAE and LBG framework is only mildly sensitive to the specific scale length of the EW distribution and the UV slope of LAEs. With a steeper EW distribution, a lower fraction of LBGs is a LAE at fixed luminosity and therefore the average \fesc is lower at fixed UV luminosity. If the limiting Ly$\alpha$ EW of LAEs is 20 {\AA} instead of 25 {\AA}, the opposite effect happens -- a higher fraction of LBGs is an LAE and the escape fraction of bright galaxies increases by a factor $\approx1.2$. A bluer UV slope implies that a fainter UV luminosity is associated to a fixed Ly$\alpha$ luminosity, leading to a similar effect. The most significant changes are seen when we lower the limiting Ly$\alpha$ luminosity of LyC-leaking LAEs (note that the average \fescs for LAEs only impacts the normalisation of the top panel in Fig. $\ref{fig:modelvar}$, and not the bottom panel). In such a faint model, the average escape fraction is highest for UV-faint galaxies with $M_{\rm UV}\approx-17$ and the average escape fraction among brighter galaxies is (by definition) a factor two lower compared to the fiducial model. This means that the majority of the ionising budget is due to significantly fainter galaxies, although a long tail exists towards brighter galaxies. Similar to the fiducial model (and any LAE emissivity model), this model has a clear peak in the UV luminosity dependence of the escape fraction.

\begin{figure}
\centering
\hspace{-1cm}    \includegraphics[width=9.2cm]{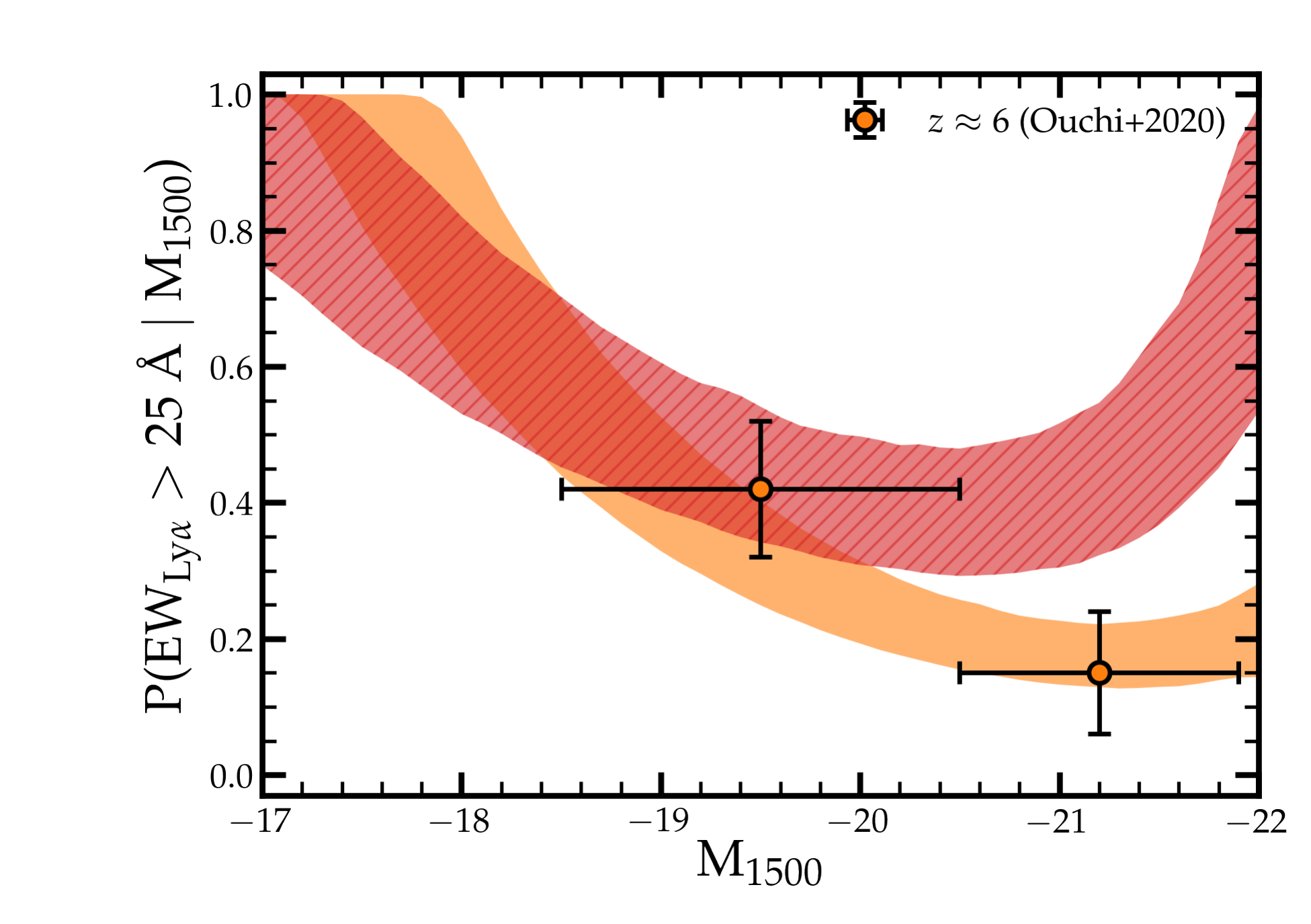}\\
\hspace{-1cm}    \includegraphics[width=9.2cm]{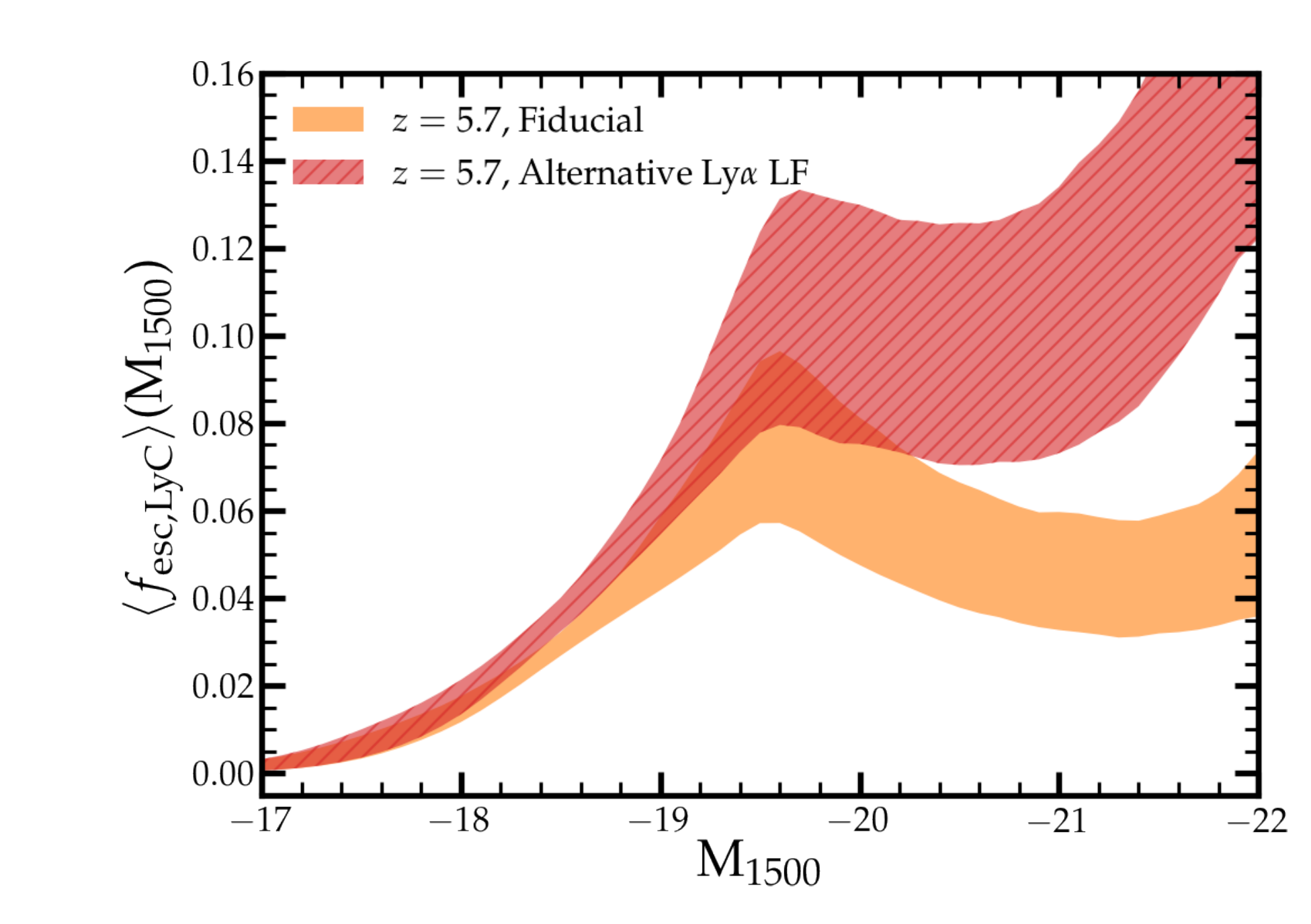}\\
\hspace{-1cm}    \includegraphics[width=9.2cm]{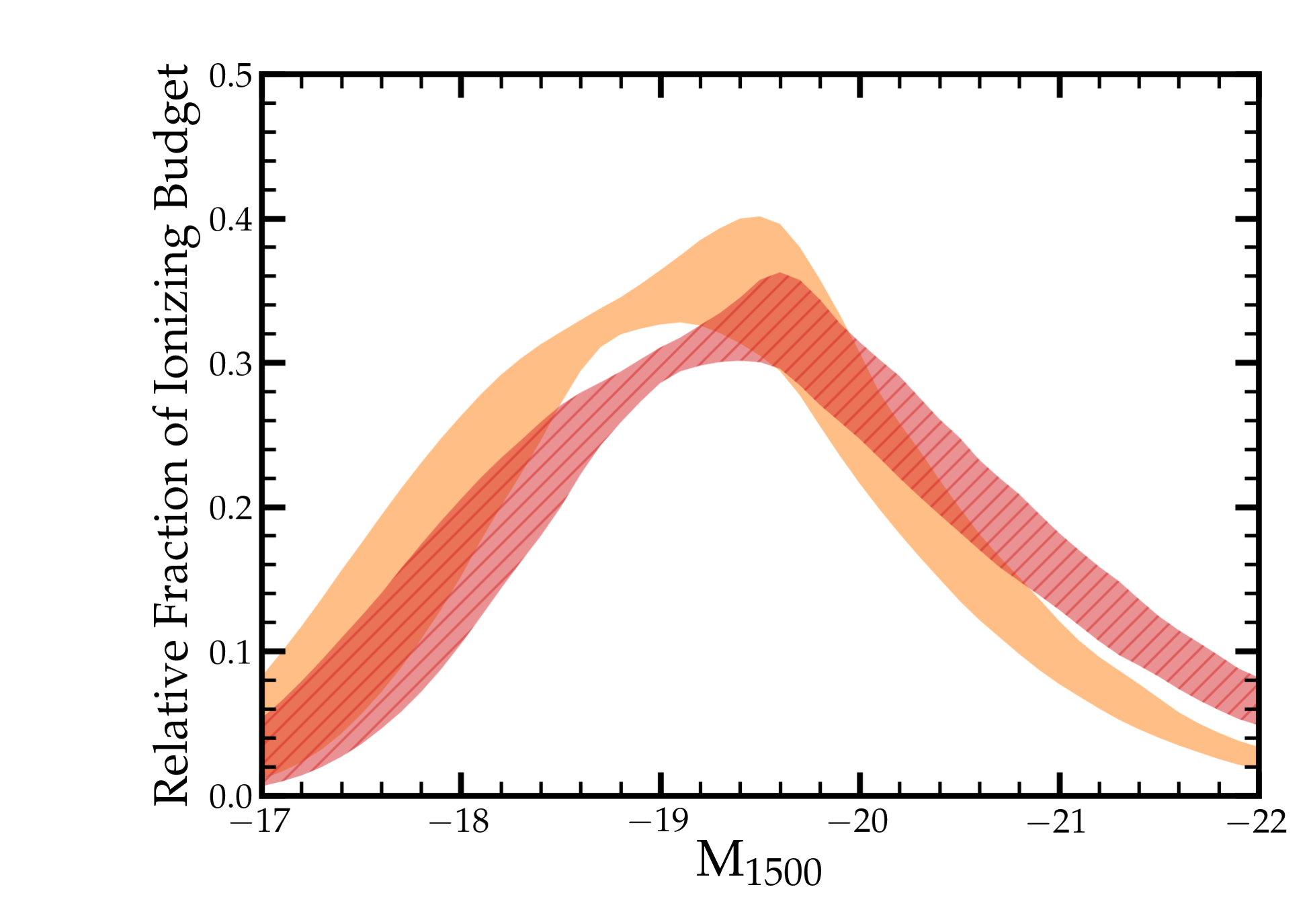}\\
\caption{The effect of varying the Ly$\alpha$ LF at $z=5.7$ (orange: \citealt{Konno18}; hatched red: \citealt{Santos16}) on the LAE fraction  (top), the average \fesc (middle) and the fractional contribution to the ionising budget as a function of UV luminosity (bottom).}
\label{fig:modelvar_2}
\end{figure}

We also investigate the impact of using the Ly$\alpha$ LF from \cite{Konno18} at $z=5.7$ (fiducial model), which shape is different from the LF measured by \cite{Santos16}. The two LFs particularly disagree on the bright-end of the LF. For other redshifts published Ly$\alpha$ LFs are in closer agreement. There is general consensus on the shape of the UV LFs over $z\approx3-7$.
As can be seen in Table $\ref{table:emissivity}$ and Fig. $\ref{fig:nion}$, the integrated Ly$\alpha$ luminosity density for these LFs are similar. However, the \cite{Santos16} LF has a higher L$^{\star}$ and a slightly shallower faint-end slope, $\alpha$. This yields a slightly different UV luminosity dependence of the LAE fraction (top panel in Fig. $\ref{fig:modelvar_2}$), which is slightly flatter flat for luminosities $M_{\rm UV}>-20$ and increases towards the brightest luminosities. The difference in the slope at fainter luminosities is due to the shallower $\alpha$ and the normalisation difference is due to the higher L$^{\star}$. Both LFs yield a consistent LAE fraction for faitner galaxies, but the \cite{Santos16} LF results in a higher LAE fraction than is observed in studies of LBGs as summarised in \citep{Ouchi2020}. A consequence is that the average \fescs in brighter galaxies increases (middle panel of Fig. $\ref{fig:modelvar_2}$) and therefore the relative contribution shifts to somewhat brighter galaxies (bottom panel).

These model variations (EW scale lengths, UV slopes, the inclusion of fainter ionisers and different Ly$\alpha$ LFs) roughly span the currently allowed edges of parameter-space, and therefore illustrate which of the results from the fiducial model are more or less certain. It is clear that the distribution of the ionizing budget - controlled by the Ly$\alpha$ luminosity limit of the faintest ioniser - is the most uncertain outcome of the LAE-emissivity model. Observations should thus prioritise testing the limiting Ly$\alpha$ luminosity of LyC leakers (as discussed in \S $\ref{sec:roadmap}$). There are also qualitative trends that are seen in all model variations, such as the existence of a peak in the relation between \fescs and UV luminosity and the relative fraction of the ionizing budget. These are therefore general outcomes of the LAE-emissivity formalism.

\section{On the possible role of AGN on the bright-end of the Ly$\alpha$ LF at $z\sim6$}  \label{sec:AGN}
The LAE emissivity framework presented in this paper does not a priori distinguish whether ionizing photons are produced by AGN or massive stars. As discussed in \S $\ref{sec:LyaLF}$, the AGN fractions among LAEs are challenging to determine at $z>5$ due to limited sensitivity of X-Ray and Radio data and inaccessibility to emission-line diagnostics \citep[e.g.][]{Calhau20}. At $z\approx3$, AGN fractions are only high at $\gg L^{\star}$ luminosities \citep{Matthee17Bootes,Calhau20}. This suggests that the AGN contribution to $\rho_{\rm{Ly\alpha}}$ is low. \cite{Matsuoka21} recently identified AGN with strong and narrow Ly$\alpha$ emission, which confirms the existence of AGN among bright LAEs at $z\approx6-7$. However, out of the five narrow-line type II AGN identified by these authors, only two have Ly$\alpha$ EW$>25$ {\AA} and would thus be picked up by traditional Ly$\alpha$ surveys that are used to construct the LAE LF. These are rare sources among high-redshift faint-AGN (2 out of 69) and their number densities are unclear. The most recent and largest spectroscopic survey of bright LAEs is the Ly$\alpha$ survey in the HEROES field \citep{Songaila18,Taylor21}. These authors report that 1/12 and 1/7 bright LAEs are AGN at $z=5.7, 6.6$, respectively. Spectroscopic surveys of slightly fainter LAEs did not report any AGN at $z\approx6-7$ \citep{Hu10,Matthee17Spec,Shibuya18}. These results imply that LAE-AGN have number densities of about $10^{-7}$ cMpc$^{-3}$, which is $\sim1000$ times lower than $\sim L^{\star}$ LAEs and suggestive of a weak AGN contribution to the emissivity.

\begin{figure}
\centering
\includegraphics[width=9.0cm]{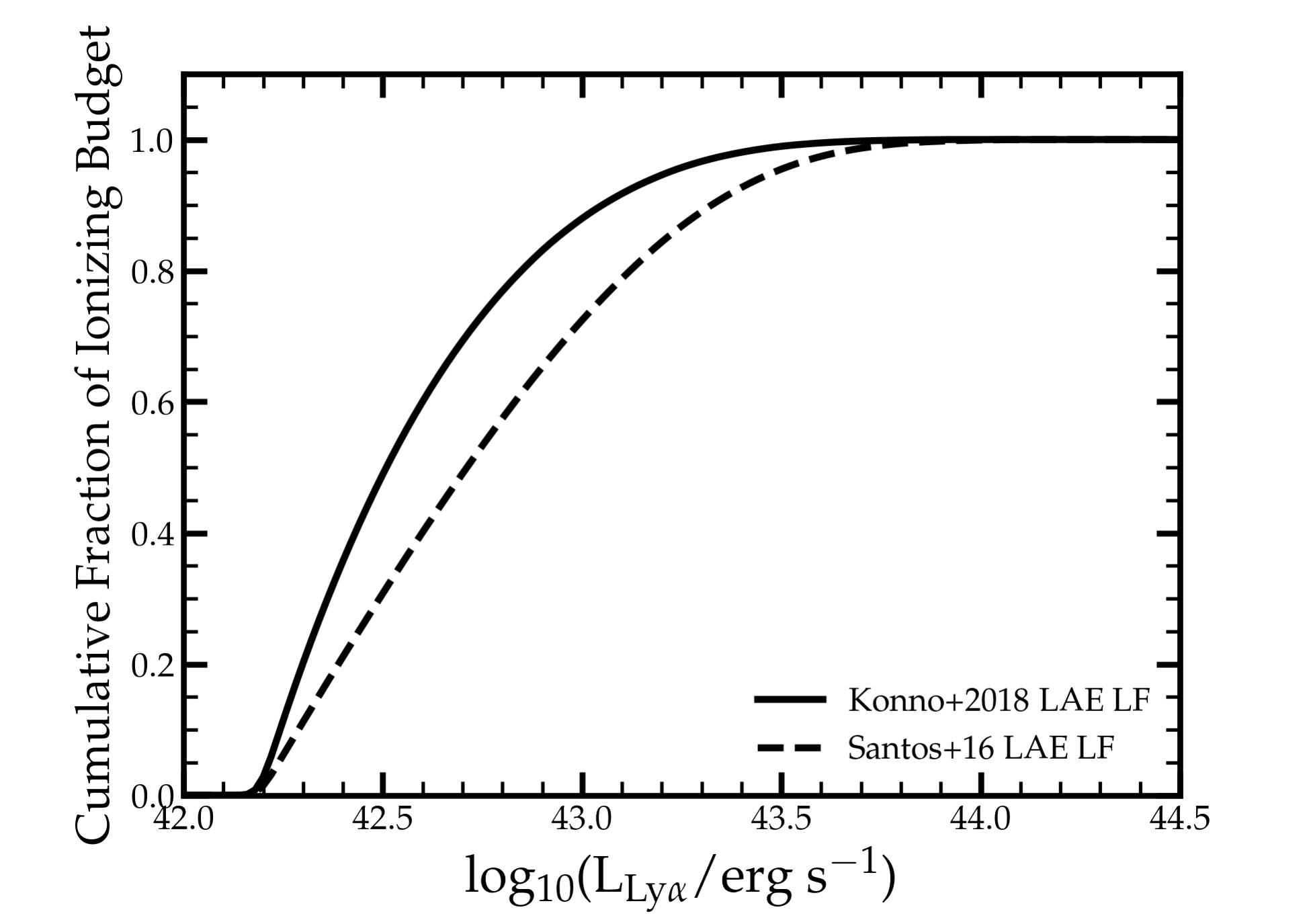}\\
\caption{The cumulative fraction of the ionizing emissivity as a function of Ly$\alpha$ luminosity in the LAE framework for two different Ly$\alpha$ LFs at $z=5.7$ (solid \citealt{Konno18}; dashed: \citealt{Santos16}).}
\label{fig:cumulative}
\end{figure}

In addition to the AGN fraction, the AGN contribution to the emissivity also depends on the shape of the bright end of the Ly$\alpha$ LF, which is contested \citep[e.g.][]{Santos16,Konno18}. Fig. $\ref{fig:cumulative}$ illustrates the relative contributions of faint and bright LAEs to the ionizing emissivity at $z=5.7$ in our fiducial framework. The LAE-AGN identified by \cite{Taylor21} and \cite{Matsuoka21} have Ly$\alpha$ luminosities above $10^{43.5}$ erg s$^{-1}$. In the most optimistic case where all LAEs that are brighter than this luminosity are AGN, AGN would still contribute $\lesssim10$ \% to the ionizing emissivity. 

Finally, we caution that both LFs shown in Fig. $\ref{fig:cumulative}$ are Schechter functions with an exponential drop-off at the bright-end. At $z\approx2-3$, various groups have reported spectroscopically-confirmed power-law components at the bright-end of the Ly$\alpha$ LF \citep[e.g.][]{Konno16,Sobral18,Spinoso20}. If such a component is present at $z\approx6$, the relative contribution of AGN could be under-estimated. As shown by \cite{Sobral18} (their Fig. 9), there are indications that the number density of this power-law component rapidly decreases above $z>3$, however it would be useful to further confirm this trend.

\section{Detailed comparison to population averaged \fescs by Pahl et al. 2021} \label{sec:Pahl}
In this section we perform a detailed investigation of the significant offset between the UV population-averaged \fescs at $z\approx3$ measured by \cite{Pahl21} and the one implied by the LAE framework.

Specifically, in the bottom-right panel of Fig. $\ref{fig:LAE_model}$ we show that our value of $\langle \xi_{\rm ion} \times f_{\rm esc}\rangle$ for galaxies with M$_{\rm UV}<-19.5$ is almost an order of magnitude lower than the measurement by \cite{Pahl21}, which is based on direct measurements from deep stacks of UV-selected galaxies with these luminosity ranges. This difference can be attributed fully to a lower average escape fraction (of $\approx1$ \% in our framework, versus 6 \% measured by \citealt{Pahl21}) that we find in this luminosity regime. This difference may partly be due to an underestimated lower fraction of LAEs in our simulation, but also due to an over-representation of LAEs in the parent sample from the study of \cite{Pahl21}. The latter may be due to the photometric selection criteria used to select LBGs (see e.g. \citealt{Kusakabe2020} for a detailed discussion). Based on the Ly$\alpha$ EW distribution (with scale length 23.5 {\AA}) and fraction of LBGs without any Ly$\alpha$ in emission (40 \%) listed by \cite{Steidel18}, we simulate the expected EW distribution for their parent sample. We reproduce the average Ly$\alpha$ EWs in the full sample and those in the four quartiles of Ly$\alpha$ EWs listed in \cite{Steidel18} to within 10 \%. In this simulated distribution, we find that the fraction of LAEs among LBGs is 21 \%, i.e. a factor $\approx3$ higher than it is in our framework based on the UV and LAE luminosity functions at $z\approx3$ (\S $\ref{sec:LBGframework}$). Correcting for this difference moves the \cite{Pahl21} measurements in the direction of our averaged escape fraction, but there is still a factor $\approx2$ difference. Additionally, we can also compare to their stack of the quartile with highest Ly$\alpha$ EW and find that 85 \% of the galaxies in that subset should be LAEs. In our model where half the LAEs have an escape fraction of $\approx50$ \%, this would imply that the average $f_{\rm esc}$ of that subset would be $0.85\times0.5\times0.5 \approx 0.21$, which is in good agreement with their measurement of $f_{\rm esc} =0.23\pm0.02$. 

\section{Author Affiliations}
\label{appendix:affiliations}
$^{3}$Kapteyn Astronomical Institute, University of Groningen, Landleven 12, 9747 AD Groningen, The Netherlands\\
$^{4}$Department of Physics \& Astronomy, Johns Hopkins University, Bloomberg Center, 3400 N. Charles St., Baltimore, MD 21218, USA\\
$^{5}$Hubble fellow\\
$^{6}$Max Planck Institut fur Astrophysik, Karl-Schwarzschild-Strasse 1, D-85748 Garching bei M\"unchen, Germany\\
$^{7}$Department of Physics, Lancaster University, Lancaster, LA1 4YB, UK\\
$^{8}$Department of Astronomy, University of Geneva, Chemin Pegasi 51, 1290 Versoix, Switzerland\\
$^{9}$Cosmic Dawn Center (DAWN), Niels Bohr Institute, University of Copenhagen, Jagtvej 128, K\o benhavn N, DK-2200, Denmark\\
$^{10}$Stockholm University, Department of Astronomy and Oskar Klein Centre for Cosmoparticle Physics,\\ AlbaNova University Centre, SE-10691, Stockholm, Sweden\\
$^{11}$Center for Gravitation, Cosmology, and Astrophysics, Department of Physics, University of Wisconsin-Milwaukee,\\3135 N. Maryland Avenue, Milwaukee, WI 53211, USA\\
$^{12}$Instituto de Investigaci\'on Multidisciplinar en Ciencia y Tecnolog\'ia, Universidad de La Serena, Ra\'ul Bitr\'an 1305, La Serena, Chile\\
$^{13}$Departamento de Astronom\'ia, Universidad de La Serena, Av. Juan Cisternas 1200 Norte,  La Serena, Chile\\
$^{14}$Department of Physics, Ulsan National Institute of Science and Technology (UNIST), Ulsan 44919, Republic of Korea\\
$^{15}$ Centro de Astrof\'{\i}sica e Gravita\c c\~ao  - CENTRA, Departamento de F\'{\i}sica, \\ Instituto Superior T\'ecnico - IST, Universidade de Lisboa - UL, Av. Rovisco Pais 1, 1049-001 Lisboa, Portugal\\
$^{16}$Instituto de Astrof\'isica de Canarias, E-38200 La Laguna, Tenerife, Spain\\
$^{17}$Departamento de Astrof\'isica, Universidad de La Laguna, E-38205 La Laguna, Tenerife, Spain\\
$^{18}$Leiden Observatory, Leiden University, PO\ Box 9513, NL-2300 RA Leiden, The Netherlands\\

\bsp	
\label{lastpage}
\end{document}